\documentclass[journal]{IEEEtran}
\usepackage{graphics} 
\usepackage{graphicx} 
\usepackage{epsfig} 
\usepackage{mathptmx} 
\usepackage{times} 
\usepackage{amsmath} 
\usepackage{amssymb}  
\usepackage{cite}
\usepackage{lipsum}
\usepackage{adjustbox}
\usepackage{multirow}
\usepackage{hyperref}
\usepackage{textcomp}
\usepackage{mathtools}
\usepackage{algorithmicx}
\usepackage[ruled]{algorithm}
\usepackage{algpseudocode}
\usepackage{xcolor}
\usepackage[justification=centering]{caption} 
\algrenewcommand\alglinenumber[1]{\tiny #1:}
\newcommand{\etal}{\textit{et al}. }

\def\sqz{\vspace{-3pt}}

\begin{document}
	\title{Deep Retinal Image Segmentation \\ with Regularization Under Geometric Priors}
	
	\author{Venkateswararao Cherukuri$^{1,2}$, Vijay Kumar B G$^{2}$, Raja Bala$^{2}$, Vishal Monga$^{1}$ 
		\thanks{$^{1}$Dept. of Electrical Engineering, The Pennsylvania State University, University Park, USA$^{2}$ Palo Alto Research Center, Palo Alto, USA.
		}%
		
	}

	
	\maketitle
	
	\begin{abstract}
		
		Vessel segmentation of retinal images is a key diagnostic capability in ophthalmology. This problem faces several challenges including low contrast, variable vessel size and thickness, and presence of interfering pathology such as micro-aneurysms and hemorrhages. Early approaches addressing this problem employed hand-crafted filters to capture vessel structures, accompanied by  morphological post-processing. More recently, deep learning techniques have been employed with significantly enhanced segmentation accuracy. We propose a novel domain enriched deep network that consists of two components: 1) a representation network that learns geometric features specific to retinal images, and 2) a custom designed computationally efficient residual task network that utilizes the features obtained from the representation layer to perform pixel-level segmentation. The representation and task networks are {\em jointly learned} for any given training set. To obtain physically meaningful and practically effective representation filters, we propose two new constraints that are inspired by expected prior structure on these filters: 1) orientation constraint that promotes geometric diversity of curvilinear features, and 2) a data adaptive noise regularizer that penalizes false positives. Multi-scale extensions are developed to enable accurate detection of thin vessels. Experiments performed on three challenging benchmark databases under a variety of training scenarios show that the proposed prior guided deep network outperforms state of the art alternatives as measured by common evaluation metrics, while being more economical in network size and inference time.

	\end{abstract}
	
	
	\IEEEpeerreviewmaketitle
	
	\section{Introduction}
	\label{sec:intro}
\begin{figure}[h]
		\begin{center}
			\includegraphics[scale=.15]{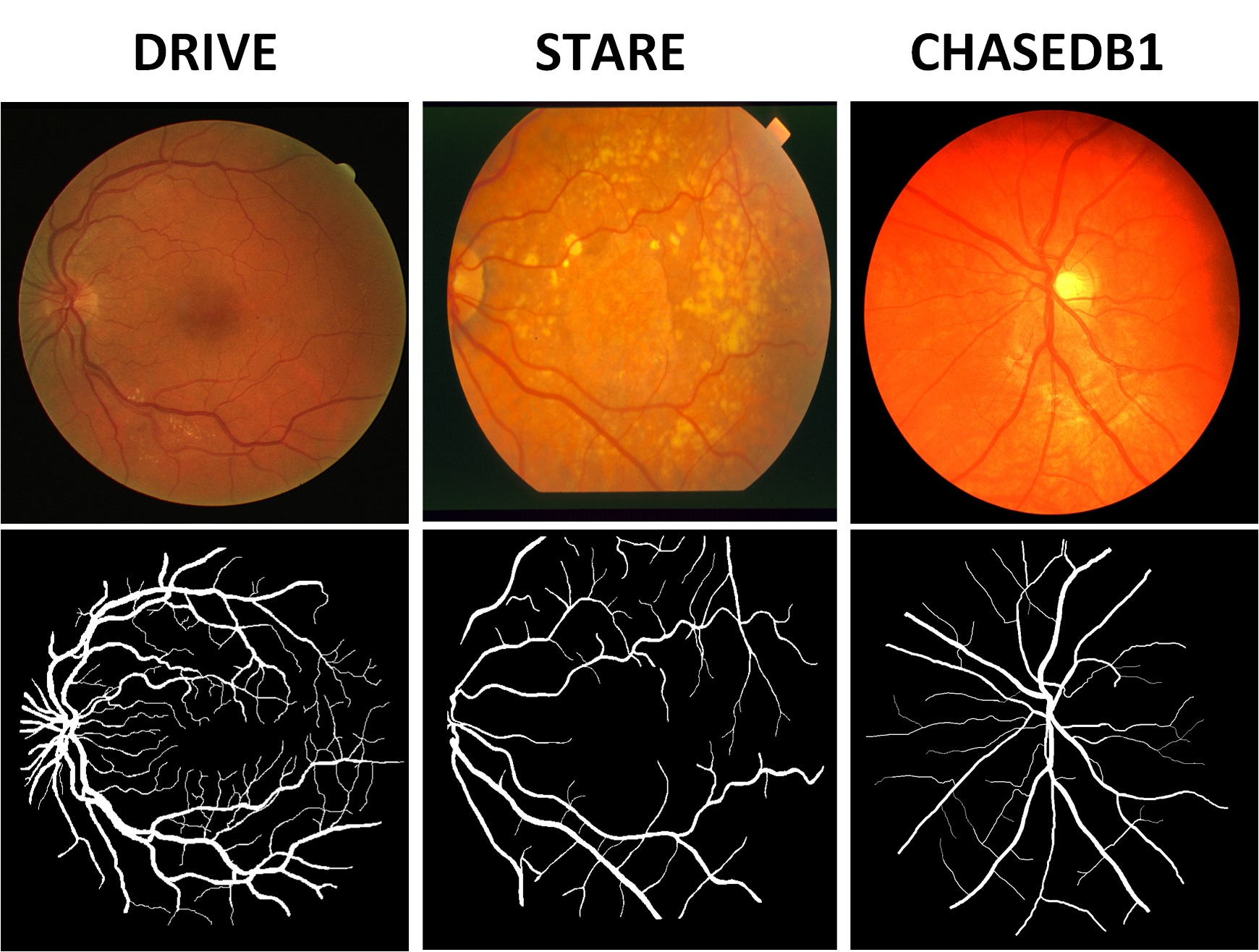}
		\end{center}
		\caption{\small{Sample images from the DRIVE, STARE, and CHASEDB1 datasets along with their ground truth (manually marked) segmentation map.}}
		\label{fig:sampImages}
	\end{figure}

	Analysis of retinal blood vessels in fundus images is crucial for diagnosis and treatment of ophthalmological diseases such as diabetic retinopathy and glaucoma. \cite{abramoff2010retinal, jelinek2009automated}. To this end, automatic vessel segmentation has gained much interest in recent years. However, this problem encounters several challenges such as inconsistencies in the shape and size of vessels, varying contrast and local intensity across different images, and interfering structures such as lesions and the optical disk. Fig. \ref{fig:sampImages} illustrates two representative example images and their corresponding manually marked vessel structures; one each from the well-known DRIVE and STARE databases \cite{staal2004ridge,hoover2000locating}.  
	
	\subsection{Related Work}
	Several methods have been developed for automatic segmentation of retinal images. These methods can be broadly classified into hand-crafted feature and learning-based methods. In the former category, hand crafted filters are designed to extract relevant features from retinal images followed by morphological image operations to segment vessels \cite{chaudhuri1989detection, wu2006adaptive, mendonca2006segmentation,hoover2000locating,zhang2010retinal,you2011segmentation,fraz2012approach,azzopardi2015trainable, zhao2015automated, yin2015vessel, zhang2016robust, wang2017enhancing, FanTIP, sheng2018retinal}. One key drawback of these methods is that the filters, once designed, are fixed and thus limited in their ability to adapt and detect a broad variety of vessel geometries.
	
	Early learning-based methods comprised extraction of hand-crafted features from the input image, which are then mapped through a pixelwise classifer such as a support vector machine (SVM) or k-Nearest Neighbor (k-NN) mapping \cite{soares2006retinal, ricci2007retinal, roychowdhury2015iterative,marin2011new, fraz2012ensemble, roychowdhury2015blood,strisciuglio2016supervised, wang2019blood, wang2019retinal}. These methods are again limited by the specific features chosen for a given dataset.

	More recently, deep learning methods have been applied to retinal vessel segmentation, wherein the features and segmentation parameters are learned simultaneously from training data. A variety of network architectures have been proposed that learn an end-end relation between training images and their corresponding manually labeled binary ground truth segmentation maps \cite{liskowski2016segmenting, li2016cross, maninis2016deep, fu2016retinal, fu2016deepvessel, son2017retinal, orlando2017discriminatively, dasgupta2017fully, yan2018three, zhang2018deep, yan2018joint, oliveira2018retinal, gu2019net}. In \cite{liskowski2016segmenting} a deep network classifies each pixel as vessel or background by constructing a patch around it. In \cite{maninis2016deep} the VGGNET architecture \cite{simonyan2014very} is fine-tuned to perform vessel segmentation and optic disc segmentation. In the work of \cite{zhang2018deep} a U-net architecture \cite{ronneberger2015u} is used for multi-label segmentation of thin and thick vessels. Yan {\em et al.} \cite{yan2018joint} propose adding a novel segment level loss function to the standard pixel-level loss to train a U-net architecture, and report increased segmentation accuracy for thin vessels. Ding {\em et al.} \cite{ding2018retinal} train a generative adversarial network to map retinal Fluorescein Angriography images to vessel maps using a novel data augmentation technique. Notably, methods based on the U-net architecture are most common \cite{son2017retinal, zhang2018deep, yan2018joint, yan2018three} and shown to deliver state of the art results, albeit that the network is often heavily parameterized.  
	
	\subsection{Motivation and Contributions}
	\label{sec:mot}
	
	While deep learning methods achieve significant improvements over their traditional counterparts, they are purely data-driven approaches that come with their shortcomings; namely their performance is strongly dependent on the quality and quantity of training data, and can degrade significantly when training samples are sparse or noisy. \color{black}Furthermore, the geometrical structure of retinal vessels is rarely exploited by existing methods. Another open challenge is accurate inference under background noise. To the best of our knowledge, previous approaches do not aim to suppress domain-induced noise that is similar in appearance to retinal vessels (i.e. the false positives). A few methods apply generic pre-processing noise suppression techniques \cite{FanTIP, liskowski2016segmenting}, which would not be entirely effective in removing domain-specific noise. Another key challenge is accurate segmentation of thin vessels. Although this problem is addressed by a few methods \cite{yan2018joint, zhang2018deep}, an improved accuracy for thin vessels often comes at the cost of slight degradation in accuracy for thick vessels.  Finally, most of the deep-learning methods employ a U-NET architecture \cite{son2017retinal, zhang2018deep, yan2018joint, yan2018three}, which is often heavily parameterized.    
	
	Keeping the aforementioned challenges in mind, \color{black}we propose a novel segmentation technique that integrates {\em a priori} knowledge of retinal vessel structures into a deep learning framework. Our network comprises two components: 1) a representation network which learns geometric (specifically curvilinear) features that are specific to retinal images, and, 2) a custom designed task network inspired by residual networks (Resnet) \cite{he2016deep} that utilizes the features obtained from the representation layer to segment vessels.  The representation layer filters are optimized jointly with the parameters of the task network. Two novel geometrical priors are imposed on the representation filters by modifying the network loss function with two regularizers: 1) an orientation diversity regularizer that encourages filters with strong responses in particular orientations spanning the $0-180$ degree range, and 2) a data adaptive noise regularizer that penalizes false positives. Further, to handle significant variations in vessel thickness, we propose a multi-scale extension of the representation layer filters.  We refer to our approach as Deep Network for Retinal Image Segmentation with Geometrical Priors (DRIS-GP).
	
	In recent work \cite{luan2018gabor}, an approach designed for image recognition (and not retinal image segmentation) proposed a method to encode orientation and scale information in a CNN. Their work modulates filters in the learning process by multiplying them elementwise with Gabor filters at varying scales and orientations. The Gabor-CNNs in \cite{luan2018gabor} and the proposed DRIS-GP however are entirely different in their analytical formulation and end goals. We perform learning of a special representation layer under problem specific constraints (the aforementioned orientation diversity and noise robustness); while no regularizers or constraints are employed in \cite{luan2018gabor}. Our learned representation filters are thus a result of constrained optimization on training data rather than Gabor or any other known analytic form.  \textbf{ Specifically our contributions\footnote{A preliminary version of this work is accepted to IEEE ICIP 2019 \cite{venkatICIP}. This paper involves substantially more analytical development in the form of: a.) new data-adaptive noise regularizer to penalize false positives, b.) a custom designed residual task network different from the U-net architecture employed in the conference paper c.) detailed derivation of back-propagation rules  d.) demonstration of computational benefits, e.) addition of a new dataset -CHASEDB1 f.) comparison with several new state of the art methods and scenarios g.) ablation studies to highlight the merits of various components of DRIS-GP.} are as follows:} 
	\begin{itemize}
		\item \textbf{Domain Enriched Deep Network:} We propose a network comprising a cascade of a domain-specific representation layer followed by a task-specific network. The representation layer is optimized under constraints germane to retinal vessel geometry. A customized residual network architecture forms the task network, utilizing features from the representation network to perform segmentation. Representation and task parameters are optimized jointly to effectively capture the curvilinear structures of blood vessels.  We argue and show that given a representation layer that is well-designed to detect curvilinear structures in the domain of retinal imagery, a residual task network needs far fewer parameters than state of the art networks tackling this problem \cite{liskowski2016segmenting,maninis2016deep,zhang2018deep,yan2018joint}, leading to significant computational benefits. 
		\item \textbf{Orientation Diversity Constraint:}  We impose a constraint on the representation filter weights that enables them to maintain desirable orientation sensitivity. This is accomplished by maximizing each given filter's response to a pattern of one orientation while minimizing its response to a pattern of orthogonal orientation. 
		\item \textbf{Data-Adaptive Noise Constraint:} To improve robustness to noise, we minimize the response of the representation layer filters to carefully selected image patches containing probable false positives. This and the orientation constraint are then added as regularizer terms to the standard reconstruction loss to obtain a novel loss function. Tractable equations are derived for the back-propagation of network parameters with respect to the loss function.
		\item \textbf{Multi-Scale Representation Network:} To handle large variations in vessel thickness, we extend the representation layer to multiple spatial scales. That is representation filters of varying sizes (one per scale) are optimized with orientation and noise constraints at multiple scales.
		
	\end{itemize}
	
	\noindent		\textit{Experimental Validation and Reproducibility:} Experimental validation is carried out on three publicly available databases: 1) DRIVE \cite{staal2004ridge}\footnote{\url{http://www.isi.uu.nl/Research/Databases/DRIVE/}}, 2) STARE \cite{hoover2000locating}\footnote{\url{http://www.ces.clemson.edu/~ahoover/stare/}}, and 3) CHASEDB1 \cite{owen2009measuring}\footnote{\url{https://blogs.kingston.ac.uk/retinal/chasedb1/}}. Extensive comparisons are performed between DRIS-GP and  state-of-the-art. For reproducibility, our model and inference code along with the results are made available at: \url{https://scholarsphere.psu.edu/concern/generic_works/mcv43nz236}. 
	
	The rest of the paper is organized as follows. A detailed description of the proposed representation layer is presented in Section \ref{sec:DRIS}.
	A custom designed residual task network is motivated and presented in Section \ref{sec:resMult}. Here we also derive back-propagation rules specifically showing the direct impact of the regularizers in learning representation layer filters and an induced (indirect) impact on the learning of task network parameters.
	A multi-scale extension is presented  in Section \ref{sec:multi-scale}. Detailed experimental validation against state-of-the-art methods is carried out in Section \ref{sec:exp}, and concluding remarks are collected in Section \ref{sec:summ}. 

	\section{Domain Enriched Representation Layer with Geometric Priors} 
	\label{sec:DRIS}
	
	\subsection{Background and Notation}
	\label{sec:not}
	\noindent 	
	We first introduce the necessary notation. Let $X \in \mathbb{R}^{M\times N}$ represent the input image, where $M$ and $N$ are the width and height of the image respectively. Let $Y\in \mathbb{R}^{M\times N}$ be the output segmented image and $Y_{g} \in \{0,1\}^{M\times N}$ be the manually labeled binary segmentation map corresponding to $X$. Let $W_{R_{k}} \in \mathbb{R}^{m\times n}$ be the $k^{th}$ convolutional filter in the geometric representation layer where $m$, $n$ represent the width, and height of the filter respectively. Similarly,  let $W_{T_{k}}^{l} \in \mathbb{R}^{m\times n\times d}$ be the $k^{th}$ convolutional filter in layer $l$ of the task network where $m$, $n$ and $d$ represent the width, height and depth of the filter respectively. Denote $b_{T_{k}}^{l} \in \mathbb{R}$ as the $k^{th}$ bias coefficient in layer $l$ of the task network. The representation and task networks are respectively $\Theta_{R} = \{W_{R_{k}}\} \forall k$ and $\Theta_{T} = \{W_{T_{k}}^{l}, b_{T_{k}}^{l}\} \forall l,k$. Finally, let the mapping functions of the representation and task networks be respectively $f$ and $F$. Then $Y = F(f(X, \Theta_{R}), \Theta_T)$. The objective of the network is to learn $\Theta_{R}$ and $\Theta_{T}$ to minimize a loss function between $Y$ and ground truth $Y_{g}$. We adopt the regression loss in light of its recent success in binary segmentation problems \cite{sironi2016multiscale}: $L_{MSE} = \frac{1}{2}\|Y-Y_{g}\|_{F}^{2}$, where $\parallel\cdot\|_{F}$ represents the Frobenius norm.
	Consistent with many recent efforts \cite{liskowski2016segmenting, li2016cross, maninis2016deep, fu2016retinal, son2017retinal, orlando2017discriminatively, dasgupta2017fully, yan2018three, zhang2018deep, yan2018joint} in this area, our proposed domain enriched deep network called DRIS-GP produces a soft (continuous) non-binary output image $Y$, which is then thresholded to obtain the final binary output.

	
	\subsection{Geometric Representation Layer}
	\label{sec:GRL}
	
	We jointly optimize representation layer filters with task network parameters to obtain representations geometrically tailored for vessel segmentation. We initially describe the framework with a single-scale representation layer and later extend it to multiple scales in Sec \ref{sec:multi-scale}. Learning the representation filters in an unconstrained manner can lead to filters that are not physically meaningful and/or practically effective. We next present two constraints based on expected prior structure of these filters. Note that these constraints are posed directly on representation layer parameters $\Theta_{R} = \{W_{R_{k}}\}$ while indirectly influencing task network parameters $\Theta_{T}$ by virtue of their joint optimization (see Section \ref{sec:backprop} for details.)

	\subsection{Orientation Diversity Constraint}
	\label{sec:orient}
	As can be observed from Fig. \ref{fig:sampImages}, retinal vessels are curvilinear structures spanning many orientations. 
	We thus develop a new regularization term that enables orientation diversity by tuning filter responses to a set of oriented image patterns. Ideally, a filter oriented in a particular given direction should yield a high response (in the sense of Frobenious norm) to an image pattern of similar orientation, and a muted response to a pattern of orthogonal orientation.  This observation guides the formulation of the orientation diversity constraint:
	\begin{equation}
	L_{Or}(\Theta_{R}) = \sum_{i=1}^{K}\big( \|W_{R_{i}}\circledast I_{O_{i}}\|_{F}^{2} - \|W_{R_{i}}\circledast I_{S_{i}}\|_{F}^{2}\big)
	\label{eq:OrientRegularizer}
	\end{equation}   
	where $K$ is the total number of filters in the geometric representation layer, $I_{S_{i}}$ is an image pattern aligned with the $i-$th orientation, $I_{O_{i}}$ is an image pattern orthogonal to the $i-$th orientation, and $W_{R_{i}}$ is the $i-$th representation filter. Note that the negative sign before $\|W_{R_{i}}\circledast I_{S_{i}}\|_{F}^{2}$ indicates that we intend to maximize this term. The oriented patterns are designed as follows:
	\begin{itemize}
		\item A 2D matrix $I$ is obtained by uniformly sampling a rectangular grid of size $I_{s}\times I_{s}$. 
		\item  $I$ is multiplied by a rotation matrix $R = \begin{bmatrix}
		\cos(\theta) & -\sin(\theta) \\ \sin(\theta) & \cos(\theta)
		\end{bmatrix}$ to obtain a set of rotated spatial pixel locations $I_{\theta}$, where $\theta$ is the desired orientation. 
		\item The desired oriented image is obtained by applying a 2D Gaussian function on $I_{\theta}$ given by:  $I_{O_{\theta}}=\exp (-(\frac{x_{\theta}^2}{c_{1}^{2}} + \frac{y_{\theta}^2}{c_{2}^{2}} ) )$
		$x_{\theta}$ and $y_{\theta}$ are the elements of $I_{\theta}$. $c_{1}$ and $c_{2}$ are chosen according to the scale of the \color{black} representation \color{black} filter. 
	\end{itemize}
	
\begin{figure}[t]
		\begin{center}
			\includegraphics[scale=.2]{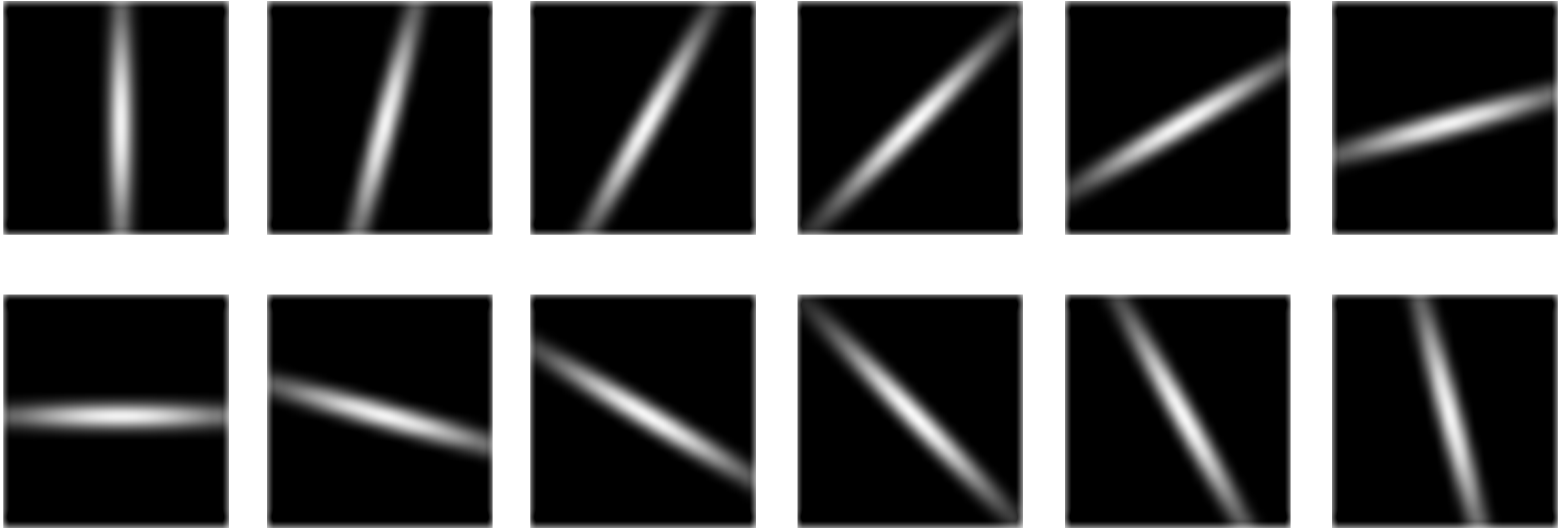}
		\end{center}
		\sqz\sqz\sqz
		\caption{\small{Synthetically generated oriented image patterns used by the orientation diversity regularizer in Eq (\ref{eq:OrientRegularizer}).}}
		\label{fig:orientImages}
	\end{figure}

	Oriented image patterns used in our work spanning the $0-180$ degree range are shown in Fig. \ref{fig:orientImages}. To design the representation filter oriented at $\theta = 0$ degrees, the first images in each row of Fig. \ref{fig:orientImages} are respectively the orthogonal and aligned image patterns\footnote{$\theta$ is measured anti-clockwise w.r.t the horizontal axis.} used in Eq (\ref{eq:OrientRegularizer}).  Unless otherwise stated, $K = 12$ representation filters (one corresponding to each orientation direction shown in Fig. \ref{fig:orientImages}) are used in our work.
	
	\subsection{Data Adaptive Noise Robustness}
	\label{sec:noiseReg}
	
	Another crucial factor that can effect the segmentation of retinal images is noise. For vessel segmentation, noise plays a major role in increasing the false-positives rate. Noise in fundus images can be of two types - 1) generated by presence of pathologies such as micro-aneurysms and hemorrhages which can be confused for vessel structures and 2) external noise generated while image acquisition \cite {walter2002contribution, abdallah2013automatic, zana1999multimodal}. It is thus desirable for the representation layer to exhibit robustness against noise components that are particularly similar to vessel structures. The most common way to address this in the literature has been to denoise the image\cite{FanTIP,ribes2014automatic,liskowski2016segmenting, rong2018surrogate} ahead of performing segmentation. Denoising methods however have the limitation of the classical trade-off: while spurious noise like components may be eliminated by denoising algorithms, they also tend to weaken some of the `true vessel structures'. Further, it is challenging and onerous to design denoising algorithms that adapt to each dataset.
	
	We take a different approach by designing representation layer filters that are naturally robust to noise. Further, we exploit the available training data to extract representative noisy patches and hence our method is data-adaptive.
	
\begin{figure}[t]
		\begin{center}
			\includegraphics[scale=.25]{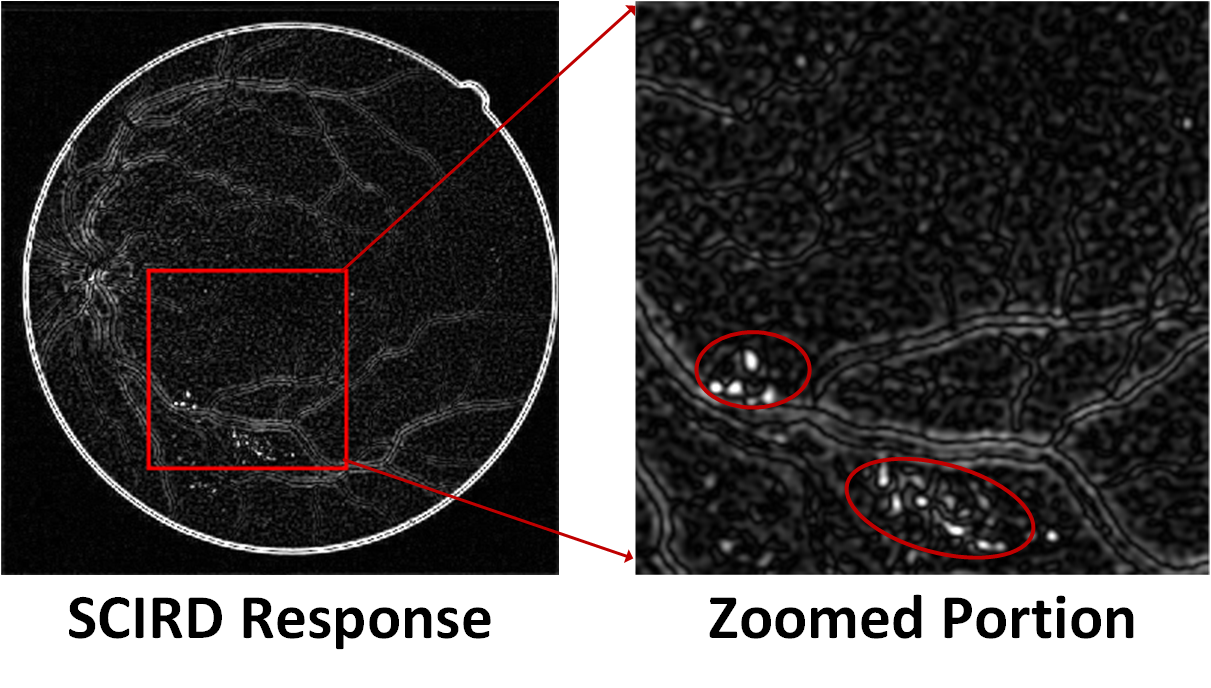}
		\end{center}
		\caption{\small{Cumulative response of SCIRD filters on an image from the DRIVE dataset. The corresponding input image is shown in Fig. \ref{fig:sampImages}. Red ellipses in the zoomed portion show potential false positives.}}
		\label{fig:noisedemo}
	\end{figure}

\begin{figure*}[!htb]
		\minipage{0.32\textwidth}
		\centering
		\includegraphics[width=\linewidth]{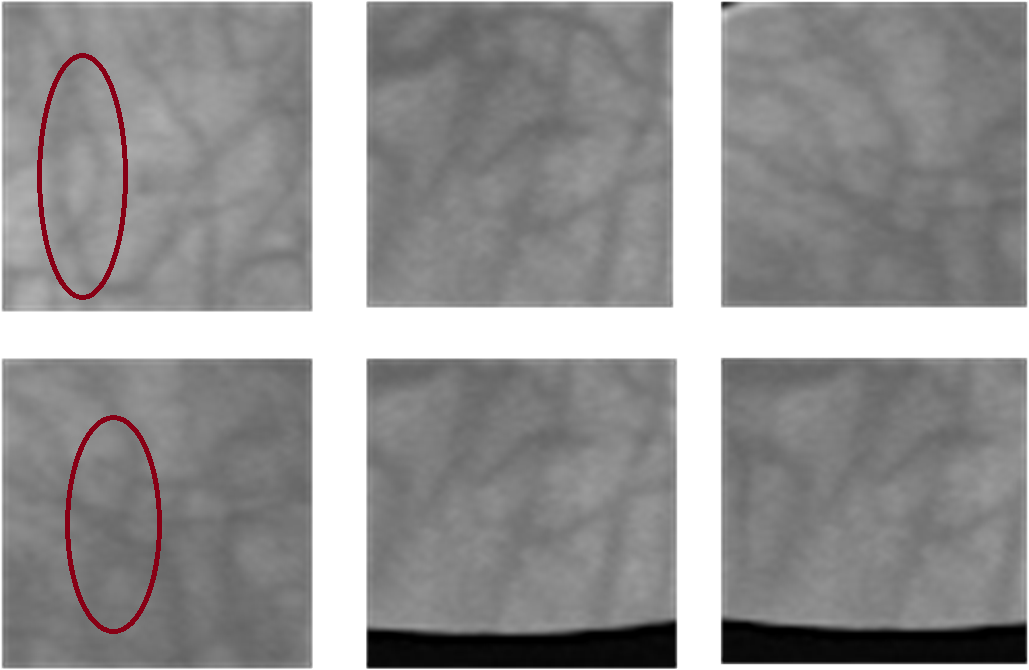}
		(a) DRIVE
		\endminipage\hfill
		\minipage{0.32\textwidth}
		\centering
		\includegraphics[width=\linewidth]{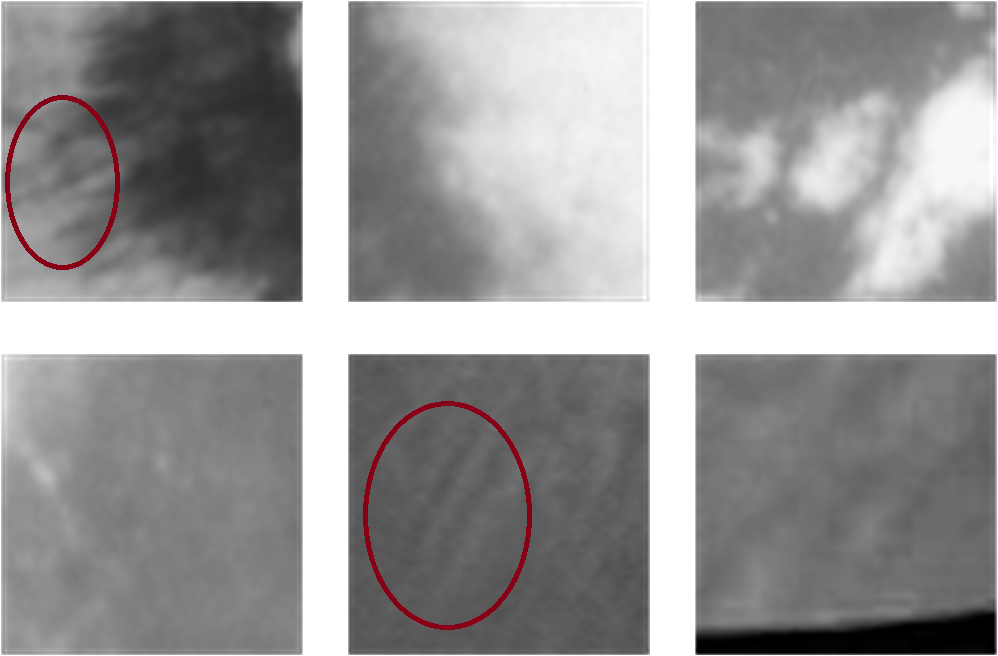}
		(b) STARE
		\endminipage\hfill
		\minipage{0.32\textwidth}%
		\centering
		\includegraphics[width=\linewidth]{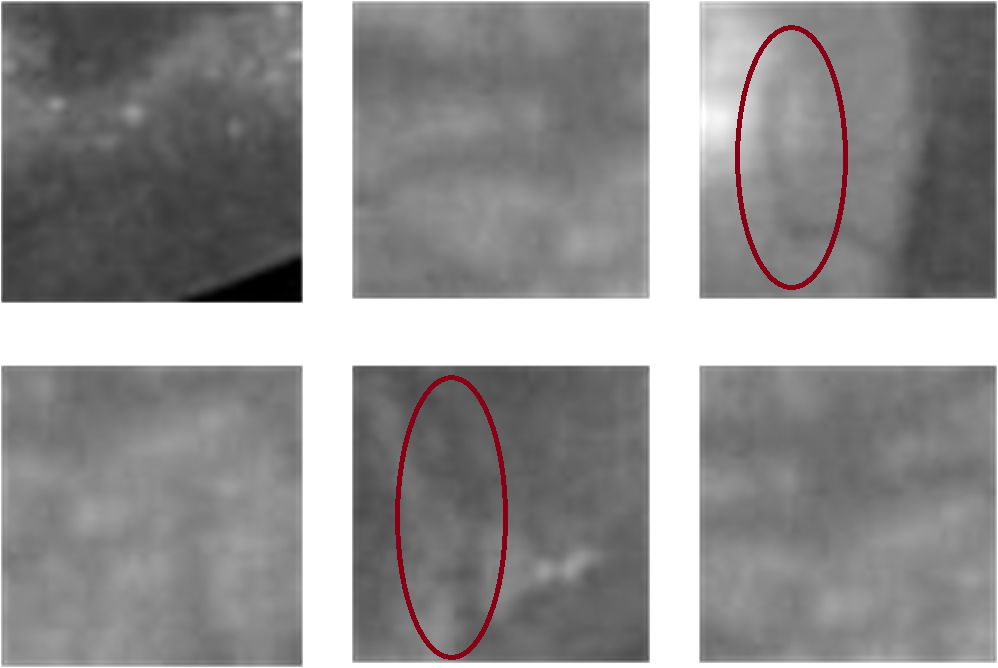}
		(c) CHASEDB1
		\endminipage
		\caption{\small{Sample noisy patches extracted from three well-known benchmark datasets. Example curvilinear structures that are {\em not} vessels are shown in red ellipses.}}
		\label{fig:noisepatches}
	\end{figure*}

	To appreciate this, we first visualize in Fig. \ref{fig:noisedemo} the averaged cumulative response of $K= 12$ Scale and Curvature Ridge Detector (SCIRD) filters \cite{annunziata2016accelerating} spanning 0-180 degrees uniformly on a retinal image. Note that we have used SCIRD filters for illustration as they have been recently shown to be state of the art for curvilinear segmentation and in particular for vessel segmentation \cite{annunziata2015scale, annunziata2016accelerating}. It can observed in Fig. \ref{fig:noisedemo} that there are several spurious components with magnitude and structure similar to that of retinal vessels, thus potentially increasing the false-positive rate for any segmentation algorithm. Because these filters are specially designed for the detection of curvilinear structures, these spurious components can be argued as a close representation of domain-specific noise. To improve robustness to such domain-specific noise, we seek to minimize the response of our optimized representation filters to noisy curvilinear structures that are often confused with vessels. Formally, we formulate and minimize the following quantity:
	\begin{equation}
	\label{eq:noisereg}
	L_{No}(\Theta_{R}) = \sum_{i=1}^{K}\sum_{j=1}^{P}\|W_{R_{i}}\circledast N_{j}\|_{F}^{2} 
	\end{equation}    
	where $N_{j}, j\in{1,\hdots, P}$ are a set of $P$ noisy image patches containing probable false-positives, extracted as follows: 
	\begin{itemize}
		\item From all training images, we extract patches of size $P_{s}\times P_{s}$ that {\em do not} contain vessel structures. 
		\item We process these patches through SCIRD filters of a particular scale (spanning 0-180 degrees), and choose the $P_{t}$ patches with highest cumulative response (in the sense of Frobenious norm). 
		\item From these top $P_{t}$ patches, we perform a visual inspection to eliminate outliers and pick $P \leq P_t$ patches.  
	\end{itemize}
	Fig. \ref{fig:noisepatches} shows representative noisy patches extracted by the procedure described above for three challenging benchmark databases. Two crucial observations can be made from this figure: 1.) It is readily apparent that these patches contain noisy curvilinear structures that can be misinterpreted as vessels, 2.) Further, it can be observed that the noise characteristics vary across the three datasets, emphasizing the data-adaptive nature of noise. Our proposed regularization term in Eq (\ref{eq:noisereg}) is indeed cognizant of this fact because it uses dataset specific training patches.
	
	The overall loss function for DRIS-GP is hence:
	\begin{align}
	\label{eq:DRIS}
	L(\Theta_T, \Theta_R) = \underbrace{\frac{1}{2}\|Y-Y_{g}\|_{F}^{2}}_{Regression\mbox{  }loss} &+ \underbrace{\alpha L_{Or}(\Theta_{R})}_{Orientation\mbox{ }Diversity}\\ \nonumber &+ \underbrace{\beta L_{No}(\Theta_{R})}_{Noise \mbox{ }Robustness}
	\end{align}
	
	where $Y = F(f(X,\Theta_R), \Theta_T)$ and $\alpha$ and $\beta$ are positive regularization constants.
	
	\noindent \textit{Remark:} We  wish to re-emphasize that the regularizers in Eqs (\ref{eq:OrientRegularizer}) and (\ref{eq:noisereg}) capture prior knowledge on representation filters without committing to any particular analytical form for the filters. Instead, these  filters are learned in a data-adaptive fashion by minimizing the cost function in Eq (\ref{eq:DRIS}). 
	
	\section{Residual Task Learning with Multi-Scale Extensions}
	\label{sec:resMult}
	\subsection{Residual Learning for Vessel Segmentation}
	\label{sec:tasknet}
	
	The majority of  recent deep-learning methods employ an U-Net architecture, inspired by \cite{ronneberger2015u}, for pixelwise segmentation problems \cite{son2017retinal, zhang2018deep, yan2018joint, yan2018three}. While such methods produce state of the art results, the designed networks involve a large number of parameters and downsampling/upsampling blocks leading to slower inference.
	We argue that since our representation layer is designed for high response to domain (i.e. curvilinear vessel) geometry, the non-linear mapping from the image to its segmentation map is significantly simplified -- a fact that can be exploited efficiently by residual learning \cite{he2016deep, nah2017deep, lim2017enhanced}.
	We therefore employ a residual network architecture \cite{he2016deep} for our task network. We note that in recent work, some residual components have been incorporated in a U-net architecture for vessel segmentation \cite{zhang2018deep}; however their overall network architecture differs {\em significantly} from ours in that the former does not exhibit a conceptual partitioning into representation and task networks (as does ours), theirs requires down/upsampling steps (ours does not) and theirs remains heavily parameterized (unlike ours).
	
\begin{figure}
		\begin{center}
			\includegraphics[scale=.75]{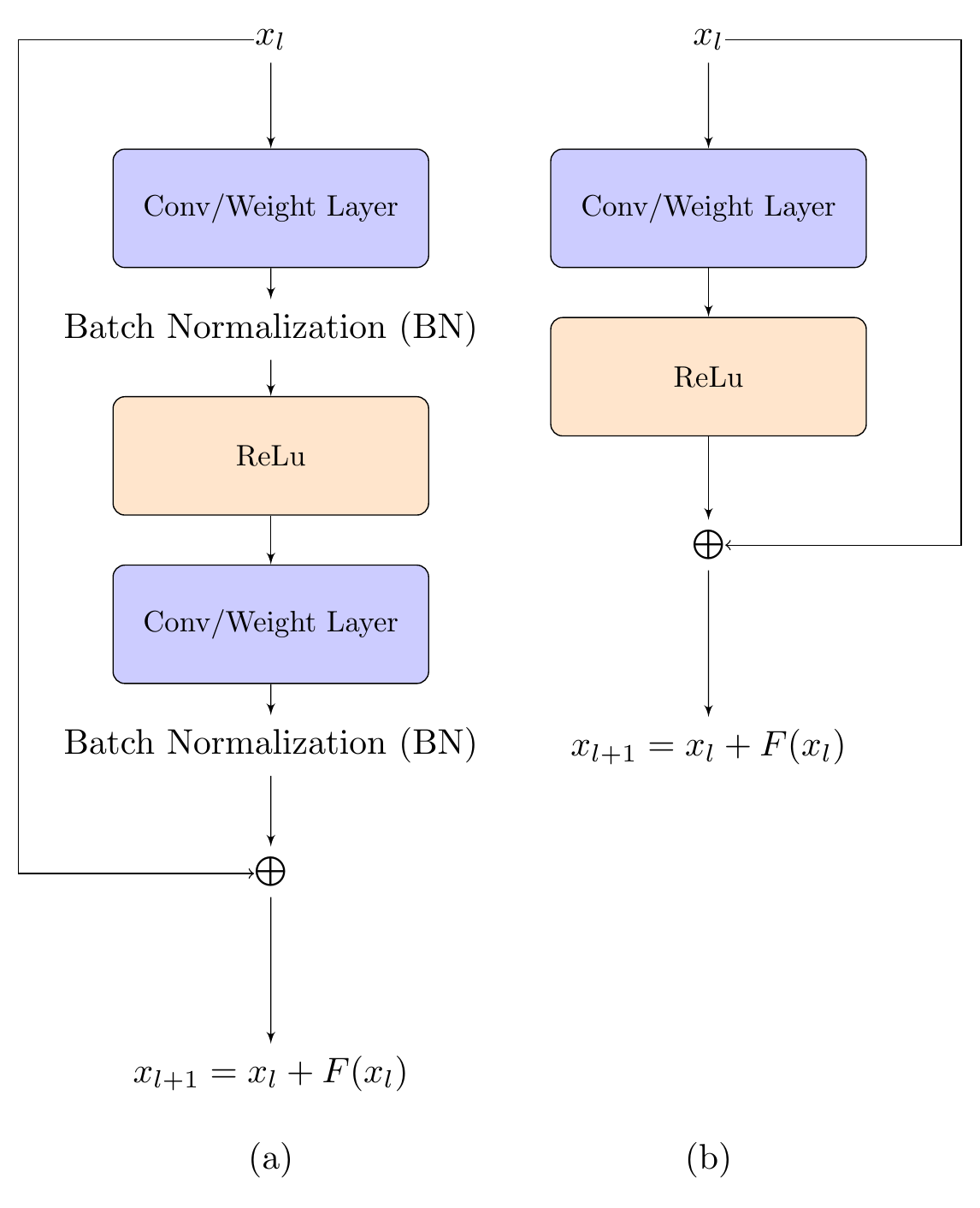}
		\end{center}
		\caption{\small{(a) A typical residual block of a resnet. (b) Simplified residual block}}
		\label{fig:resNet}
	\end{figure}

	Fig. \ref{fig:resNet}(a) shows the standard building block for common residual networks. As can be observed, the output of each residual block $x_{l+1}$ learns incremental features on top of the features learned in previous layer. Mathematically, the output of layer $x_{l+1}$ is given by:
	\begin{equation}
	\label{eq:res_eq}
	x_{l+1} = x_{l} + F(x_{l}) = g(x_{l})
	\end{equation}
	
	where $g()$ is the functional mapping to be learned and $F()$ is the residual mapping given that the output of the previous layer is added.

	To understand the benefits of residual learning for our problem, consider the case when $g(x_{l})$ is an identity function. In this case, a residual block has to learn a mapping to an all-zero output, which is considerably simpler. More generally, simplified non-linear mappings can be designed with a much more economical network with residual learning \cite{lim2017enhanced}. Hence, it is apparent that if rich problem specific features are obtained in the early representation layers, the learning of a residual task network simplifies further. 
	
	Note that our task network benefits from highly informative vessel features extracted by optimizing representation layer filters with domain-specific regularization as in Section \ref{sec:DRIS}. We thus further simplify the design of our residual network via two changes - see Fig. \ref{fig:resNet}(b). First, the Batch-Normalization (BN) layers are removed. It has been argued recently that normalizing features gets rid of the range flexibility which affects performance of the network \cite{lim2017enhanced, nah2017deep}. Further, since the BN layer occupies the same memory as the output of the previous convolutional layer, a GPU can efficiently utilize the extra memory afforded by removal of the BN layers. Secondly, we design our residual block with only a single convolutional layer in contrast to a typical residual block containing two convolutional layers. This luxury is again afforded by the efficacy of our representation layer. That is, the input to the task network already captures vessel-like structures to a first order, thus simplifying residual mappings $F()$ that must be learned. For the same number of parameters (against a standard double convolutional layer set-up) our proposed resnet can hence deploy more residual blocks allowing for incremental features to be more effectively learned towards approximating the ground truth segmentation map.
	Specific design details and methodology for selecting task network parameters are given in Section \ref{sec:tasknet_details}.    
	
\begin{figure*}[t]
		\begin{center}
			\includegraphics[scale=1.4]{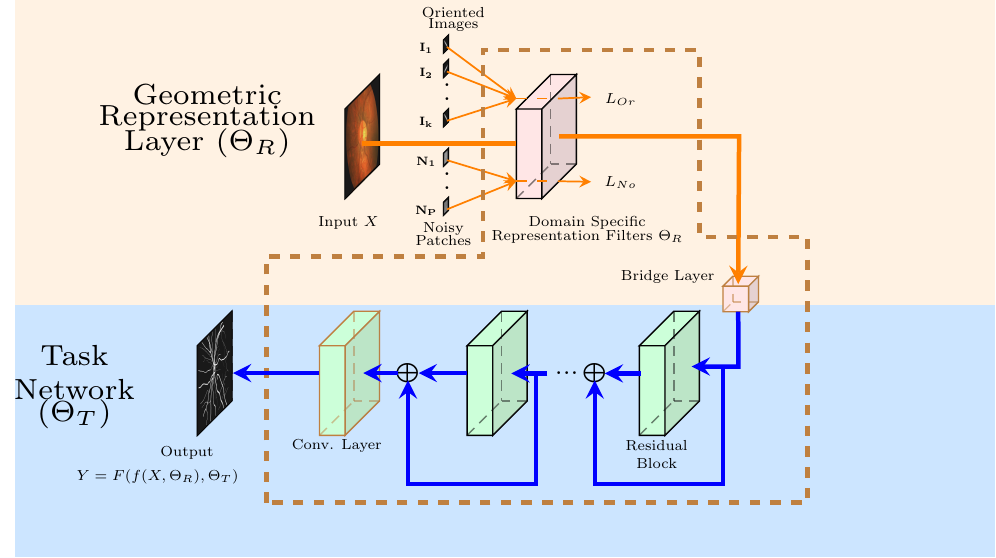}
		\end{center}
		\caption{\small{DRIS-GP Network Architecture: The representation network comprises a set of geometrically constrained filters. The output of the representation layer is connected to the task network (resnet) via a bridge layer that has $D$ filters of size $1\times 1 \times K$. The task network performs the segmentation. Both representation and task network filters (included inside the brown dashed lines) are jointly learned.}}
		\label{fig:Archi}
	\end{figure*}

	\textbf{Network Architecture:} The overall DRIS-GP architecture is shown in Fig. \ref{fig:Archi}. The representation and task network are connected by a bridge convolutional layer that comprises $D$ filters of size $1\times 1\times K$. $K$ depends on the number of filters used in the representation layer. Implementation details of the representation layer are found in Section \ref{sec:multi-scale}. The output of the task network is sent through a $3\times 3\times D$ convolutional layer to obtain the final output segmentation map $Y$. 
	
	\subsection{Back-propagation: Influence of Priors on Representation and Task Network Parameters}
	\label{sec:backprop}
	We learn $\Theta_T$, $\Theta_R$ by minimizing $L(\Theta_T, \Theta_R)$ in Eq (\ref{eq:DRIS}) using a stochastic gradient descent method \cite{lecun1998gradient, werbos1994roots}. Note that both the regularization terms are differentiable with respect to the network parameters, thus facilitating tractable learning. In particular, weights are updated by the following equation: $\Theta^{t+1} = \Theta^{t} - \eta\frac{\partial L}{\partial\Theta^{t}}$, where, $\Theta = \{\Theta_R, \Theta_T\}$, $\eta$ represents the learning rate, and $\Theta^{t}$ represents the values of weights at iteration $t$. Let $w_{R_{m,n}}$ be the $(m,n)^{th}$ coefficient of geometric representation layer filter $W_{R_{k}}$ and $w_{T_{m,n}}^{l}$ be the $(m,n)^{th}$ coefficient of task network filter $W_{T_{k}}^{l}$. The equation for computing the gradient with respect to $w_{T_{m,n}}$ is given by:
	\begin{equation}\label{eqwtk}
	\frac{\partial L}{\partial w_{T_{m,n}}^{l}} = -(Y_{g} - Y)\diamond\frac{\partial Y}{\partial w_{T_{m,n}}^{l}} 
	\end{equation}
	where $\diamond$ between two matrices $A$ and $B$ is defined as $\sum_{i,j}A_{i,j}B_{i,j}$. Note that the task network parameter $w_{T_{m,n}}^{l}$ does not depend directly on $L_{Or}$ and $L_{No}$, hence these terms are not reflected in the gradient computation. However, as $Y$ is dependent on the representation layer parameters, the task network parameters are indirectly influenced by $L_{Or}$ and $L_{No}$. $\frac{\partial Y}{\partial w_{T_{m,n}}^{l}}$ is obtained by standard back-propagation rule \cite{lecun1998gradient, werbos1994roots}. The gradient with respect to $w_{R_{m,n}}$ is given by:
	\begin{equation}\label{eq:repwk}
	\frac{\partial L}{\partial w_{R_{m,n}}} =  -(Y_{g} - Y)\diamond\frac{\partial Y}{\partial w_{R_{m,n}}}  + \alpha \frac{\partial L_{Or}}{\partial w_{R_{m,n}}} + \beta \frac{\partial L_{No}}{\partial w_{R_{m,n}}}
	\end{equation} 
	$\frac{\partial Y}{\partial w_{R_{m,n}}}$ can be computed using the standard back-propagation rule. $\frac{\partial L_{Or}}{\partial w_{R_{m,n}}}$ is given by:
	\begin{equation}
	\label{eq:lor}
	\frac{\partial L_{Or}}{\partial w_{R_{m,n}}} = 2 (W_{R_{k}}\circledast I_{O_{k}})\diamond I_{O_{k}}' - 2 (W_{R_{k}}\circledast I_{S_{k}})\diamond I_{S_{k}}'
	\end{equation}
	where $I_{O_{k}}' = [I_{O_{k_{i-m,j-n}}}]$, $I_{O_{k_{i-m,j-n}}}$ is the $(i-m,j-n)^{th}$ element of $I_{O_{k}}$. $I_{S_{k}}'$ is defined in a similar way. Similarly, $\frac{\partial L_{No}}{\partial w_{R_{m,n}}}$ is given by:
	\begin{equation}
	\label{eq:lor}
	\frac{\partial L_{No}}{\partial w_{R_{m,n}}} = 2 \sum_{j=1}^{P}(W_{R_{k}}\circledast N_{j})\diamond N_{j}'
	\end{equation}
	Where $N_{j}'$ is defined similar to $I_{O_{k}}'$. It can be observed that both the regularization terms $L_{Or}$ and $L_{No}$ have direct influence on $w_{R_{m,n}}$ which indirectly affects $w_{T_{m,n}}$. 
	
	\subsection{Multi-Scale Representation Layer: Handling Varying Vessel Thickness and Enhance Thin Vessel Segmentation}
	\label{sec:multi-scale}
	
	One of the key practical challenges in segmenting vessels is their {\em thickness inconsistency}. Since the majority of the vessels are thick in nature, often thin vessels are poorly segmented. Efforts have been made via deep learning approaches to address this problem. In \cite{zhang2018deep, yan2018joint}, the ground-truth vessel labels are further divided into thin and thick vessels to learn a multi-class segmentation network (as opposed to the standard binary segmentation). Similarly in \cite{yan2018joint}, a new loss term is introduced to place greater emphasis on thin vessels. 
	
\begin{figure}[t]
		\begin{center}
			\includegraphics[scale=1.5]{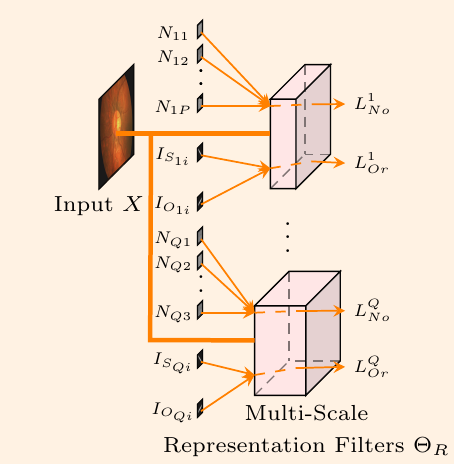}
		\end{center}
		\caption{\small{Illustration of the multi-scale representation layer.}}
		\label{fig:mult-scale}
	\end{figure}

	Departing from the aforementioned approaches, we handle vessel thickness diversity by optimizing the representation filters at multiple scales, i.e.\ filters of different spatial sizes are picked for each scale -- to build a multiscale extension of  DRIS-GP (which we will henceforth call MS-DRIS-GP).  The orientation patterns and noise patches are generated at each scale with the same procedures described earlier. Likewise, the regularizers in Section \ref{sec:DRIS} are applied per scale.  Fig. \ref{fig:mult-scale} illustrates the multi-scale representation layer. The new loss function with such a layer is given by:  
	\begin{equation}
	\label{eq:MSDRIS}
	L^{MS}(\Theta_T, \Theta_R) = \frac{1}{2}\|Y-Y_{g}\|_{F}^{2} + \alpha L_{Or}^{MS} + \beta L_{No}^{MS}
	\end{equation}
	where $L_{Or}^{MS}$ and $ L_{Or}^{No}$ are defined as:
	\begin{equation}
	L_{Or}^{MS} = \sum_{s=1}^{Q}\sum_{i=1}^{K}\|W_{R_{si}}\circledast I_{O_{si}}\|_{F}^{2} - \|W_{R_{si}}\circledast I_{S_{si}}\|_{F}^{2}
	\end{equation}
	
	\begin{equation}
	L_{No}^{MS} = \sum_{s=1}^{Q}\sum_{i=1}^{K}\sum_{j=1}^{P}\|W_{R_{si}}\circledast N_{sj}\|_{F}^{2} 
	\end{equation}
	where $Q$ is the number of scales, $W_{R_{si}}$ represents the $i^{th}$  filter for scale $s$. $N_{sj}$ represents the $j^{th}$ noisy patch for scale $s$, $I_{O_{si}}$ and $I_{S_{si}}$ are multi-scale extensions of $I_{O_{i}}$, $I_{S_{i}}$. 
	
	\noindent \textbf{Implementation Specifics for Multi-scale Representation Network:} Our representation layer comprises of 5 convolutional filter banks with each filter bank aimed at extracting features pertaining to different scales labeled as scale 1-5. Each set of representation filters pertaining to a particular scale comprises of 12 filters that are intended to span 0-180 degrees uniformly. The sizes of filters employed for each scale in increasing order are set to: $3\times 3$, $5\times 5$, $7\times 7$, $9\times 9$ and $11\times 11$. The values of $I_{s}$ used for constructing oriented images for different scales in increasing order are given by: 6, 10, 14, 18, and 22. The values of Gaussian parameter $c_{1}$ in the same order are given by: 1, 2, 3, 4, and 5. $c_{2}$ for all cases is chosen to be 10. For the experiments related to the single scale representation network, we adopt the configuration described for scale 3. We use these parameter values for remainder of the paper, unless otherwise mentioned.     
	
	\subsection{Cross-validation for Task-Network Parameters}
	\label{sec:tasknet_details}
	
	Two parameters primarily guide the design of a resnet: the number of layers $L$ and the depth $D$ of the convolutional filters used in each residual block. Typically $D$ $3\times 3\times D$ convolutional filters with $D \geq 64$ are used. We build our task network with $L=14$ and $D=32$ differing from the standard configuration of $D \geq 64$. These choices are arrived at by a cross-validation approach.  Table \ref{tab:results_task_filters} shows the F1 score (F1 -- as defined in Section \ref{sec:expsetup}) of our DRIS-GP (at scale 3) with varying filter depth $D$ on the DRIVE dataset with train-test configuration as described in Section \ref{sec:specConfig}. As can be observed, there are no noticeable gains in accuracy for $D \geq 32$. Further, the performance degrades for $D = 128$ and $256$ due to over fitting. Similarly, Table \ref{tab:results_task_layers} reports the accuracy for different values of $L$. Diminishing returns are observed for $L \geq 14$. For the rest of the paper, we hence fix $L=14$ and $D = 32$. Finally, while the DRIVE dataset is used to tune these parameters, we found these choices to hold for both STARE and CHASE databases, and for both single- and multi-scale versions of DRIS-GP. 
	
	\begin{table}[t]
		\caption{\small{DRIS-GP Performance with fixed $L = 14$ and varying $D$}}
		\label{tab:results_task_filters}
		\begin{center}
			\begin{tabular}{|l|l|l|l|l|}
				\hline
				Number of Filters & $D=32$ &  $D=64$ & $D=128$ & $D=256$\\
				\hline
				F1 & $0.8208$ & $0.8211$  & $0.8206$ & $0.8201$\\
				\hline
				
			\end{tabular}
		\end{center}
	\end{table}
	
	\begin{table}[t]
		\caption{\small{DRIS-GP Performance with $D = 32$ and varying $L$}}
		\label{tab:results_task_layers}
		\begin{center}
			\begin{tabular}{|l|l|l|l|l|l|}
				\hline
				Number of Layers & $L=10$ &  $L=12$ & $L=14$ & $L=16$& $L=18$\\
				\hline
				F1 & $0.8182$ & $0.8197$  & $0.8208$ & $0.8210$ & $0.8212$\\
				\hline

			\end{tabular}
		\end{center}
	\end{table}
	
\color{black}
\subsection{Validation of Task Network Architecture}
\label{sec:abl_tasknet}
In Section \ref{sec:tasknet}, we proposed a simplified residual architecture by leveraging the fact that the representation network is specially tuned for responding to curvilinear vessel structures. We now validate the benefits of the proposed network over the standard residual architecture, named ST-RES for brevity. In the first experiment, we compare the design of our network with ST-RES comprising a batch-normalization layer and two convolutional layers in each residual block. To be consistent with the number of parameters we use $L = 7$ in  ST-RES. The results are reported in Table \ref{tab:residual_abl} on the DRIVE dataset with the train-test configuration described in Section \ref{sec:specConfig}. It is readily seen that DRIS-GP outperforms ST-RES. We believe this is attributed to the fact that DRIS-GP contains double the number of residual blocks for the same number of parameters. We also note that the inference time of ST-RES is greater than that of DRIS-GP. This is due to the fact that the batch-normalization layers in ST-RES occupy the same memory as the output of the previous convolutional layer, hence decreasing GPU efficiency. In the second experiment, we derive a variant of ST-RES by using a single convolutional layer in the residual block and retaining the Batch-Normalization (BN) layers. We call this configuration BN-RES. Note that the two networks have the same number of residual blocks. As seen in Table \ref{tab:residual_BN}, the difference between BN-RES and DRIS-GP is marginal in terms of F1 score; however DRIS-GP has superior inference time. The marginal decrease can be attributed to the fact that the BN layers, which normalize the output features at each layer, may not play an active role if effective features are already extracted in the previous layers. Overall, we argue that our proposed residual architecture exploits the efficacy of the representation layer, and is more computationally efficient than the standard residual architecture.       

	\begin{table}[t]
	\color{black}
	\caption{\small{DRIS-GP Performance with standard residual architecture and our proposed DRIS-GP}}
	\label{tab:residual_abl}
	\begin{center}
		\begin{tabular}{|l|l|l|}
			\hline
			Task-Net & ST-RES &  DRIS-GP\\
			\hline
			F1 & $0.8193$ & $0.8208$\\
			\hline
			Inf. Time  & 31ms & 26ms \\
			\hline

		\end{tabular}
	\end{center}

\end{table}

	\begin{table}[t]
	\color{black}
	\caption{\small{DRIS-GP Performance compared with BN-RES}}
	\label{tab:residual_BN}
	\begin{center}
		\begin{tabular}{|l|l|l|}
			\hline
			Task-Net & BN-RES &  DRIS-GP\\
			\hline
			F1 & $0.8206$ & $0.8208$\\
			\hline
			Inf. Time  & 33ms & 26ms \\
			\hline

		\end{tabular}
	\end{center}
	
\end{table}
\subsection{Cross Entropy Vs. Regression Loss}
\label{sec:abl_loss}
In this section we examine via an ablation study the effect of the choice of loss function on network performance, comparing our choice of regression loss function (Eq. \ref{eq:DRIS}) with the commonly used cross-entropy loss \cite{liskowski2016segmenting, maninis2016deep, zhang2018deep, yan2018joint, yan2018three}. Results are reported in Table \ref{tab:diffloss} on the DRIVE dataset with train-test configuration as described in Section \ref{sec:specConfig} and at scale 3. As can be observed, noticeable gains are not obtained by using a cross-entropy loss function.  \color{black}Further, in order to train with cross-entropy loss, the output of the task network is sent to a $3\times 3\times D\times 2$ convolutional layer, while training with regression loss calls for a $3\times 3\times D\times 1$ layer. Cross-entropy loss thereby adds $3\times 3\times D\times 1$ more network parameters compared to regression loss. Hence we elect the regression loss for all remaining experiments. \color{black}     

	\begin{table}[t]
		\color{black}
	\caption{\small{DRIS-GP Performance with cross-entropy loss and regression loss}}
	\label{tab:diffloss}
	\begin{center}
		\begin{tabular}{|l|l|l|}
			\hline
			loss & Regression &  cross-entropy\\
			\hline
			F1 & $0.8208$ & $0.8211$\\
			\hline

		\end{tabular}
	\end{center}
\color{black}
\end{table}
\color{black}
	
	\section{Experimental Evaluation}
	\label{sec:exp}
	\subsection{Experimental Setup}
	\label{sec:expsetup}
	\noindent \textbf{Datasets, Training and Test Setup:} We evaluate DRIS-GP on three standard datasets widely used for this problem. The first dataset called DRIVE \cite{staal2004ridge} contains 40 fundus images with manually labeled ground truth maps. We randomly choose 20 images for training and use the remaining images for testing. To remove selection bias, we repeat the experiment five times and report the averaged result. The second dataset called STARE \cite{hoover2000locating} contains 20 manually labeled fundus images. We report average results on 5 random selections of 10 training images and 10 test images.  The third dataset,  CHASEDB1 \cite{owen2009measuring}, contains 28 manually labeled fundus images. We report averaged results on 5 random selections of 14 training images and 14 test images.
	
	\noindent \textbf{Training Patch Extraction and Parameter Selection:} As is common in segmentation problems \cite{liskowski2016segmenting, roychowdhury2015blood, fraz2012ensemble, yan2018joint,yan2018three}, in order to obtain a sizable training set, we extract patches of size $128\times 128$ from training images with their corresponding ground truth. Patches are extracted via the procedure in \cite{yan2018joint}. Final inference is carried out on the entire image. Consistent with other approaches \cite{roychowdhury2015blood, fraz2012ensemble, yan2018joint, yan2018three}, we extract patches from the green channel of RGB images.  Approximately 7000 training patches are extracted for DRIVE and CHASEDB1, while approximately 4000 patches are extracted for STARE. As described in Section \ref{sec:noiseReg}, $P = 100$ training noisy patches of patch size $P_{s} = 64$ are extracted, and oriented images described in Section \ref{sec:orient} are synthetically generated, and used for network learning with noise and orientation regularizers in Eq (\ref{eq:DRIS}).  The regularization weights $\alpha$ and $\beta$ in Eq. (\ref{eq:MSDRIS}) are chosen as $10e^{-7}$ and $10e^{-5}$ using a nested cross-validation procedure \cite{cawley2010over, monga2017handbook}. Since the (MS)-DRIS-GP output is soft, we use a threshold consistent with  existing work \cite{li2016cross, yan2018joint, yan2018three} to obtain the binary output.
	For optimization, we employ the Adam Optimizer \cite{kingma2014adam} with a learning rate of $5\times 10^{-4}$, batch size of $64$ and number of epochs $=60$. All our experiments are performed on an NVIDIA Titan X GPU (12GB) with the TensorFlow package \cite{abadi2016tensorflow}.  
	
	\noindent \textbf{Evaluation Metrics:} Standard metrics including Area under ROC curve (AUC), Accuracy (Acc), Sensitivity (Sens), Specificity (Spec), and Dice-Overlap/F1-score (F1) are used for evaluating DRIS-GP. Additionally, we report precision-recall curves for selected experiments. The mathematical definitions of selected metrics are given by: 
	\begin{equation*}
	F1 = \frac{2TP}{2TP+FP+FN}\mbox{,  }Acc = \frac{TP + TN}{TP + TN + FP + FN}
	\end{equation*} 
	\begin{equation*}
	Sens = \frac{TP}{TP+FN}\mbox{,  }Spec = \frac{TN}{TN + FP}
	\end{equation*}
	where $TP$, $TN$, $FP$, and $FN$ corresponds to true positives, true negatives, false positives, and false negatives respectively. \color{black}Performance is evaluated for the pixels inside the Field of View (FOV) for the respective datasets. For DRIVE, we used the FOVs that came with the dataset; while for STARE and CHASEDB1, FOVs were not publicly available, hence we employed the FOVs used in \cite{liskowski2016segmenting}. Note that FOVs only reduce the number of TNs typically found in background regions, and hence do not alter F1 and sensitivity scores for all practical purposes. 

	\noindent	\color{black} \textbf{Initialization of Network Parameters} 
	\label{sec:net_init}
	In numerical optimization of network parameters, an initial seed point must be provided. Most popular deep learning frameworks for this problem \cite{zhang2018deep, yan2018joint, liskowski2016segmenting} use the Xavier initialization \cite{glorot2010understanding}. We therefore also adopt Xavier initialization for our task network parameters. Because the representation layer aims to capture curvilinear structures at different orientations, an intelligent initialization can facilitate faster convergence. We initialize our representation layer in $3$ different ways: 1) Xavier initialization; 2) synthetic orientation patterns described in section \ref{sec:orient}; 
	and 3) SCIRD filters \cite{annunziata2016accelerating} -- state of the art oriented filters for curvilinear feature detection. 
	The results in the form of F1 score are reported in Table \ref{tab:init} on the DRIVE dataset for the train-test configuration described in \ref{sec:specConfig}. It can be observed that our network can be used with a variety of initializers, the exact choice depending partially on the dataset(s). Unless otherwise stated, the representation layer in our proposed (MS)-DRIS-GP is initialized with SCIRD filters, as we found that this choice provides the fastest convergence in the training phase.
	\color{black}
	\begin{table}[t]
		\caption{\small{DRIS-GP Performance with different initializations for representation layer}}
		\label{tab:init}
		\begin{center}
			\begin{tabular}{|l|l|l|l|l|}
				\hline
				Initialization & Xavier &  Synthetic orientation patterns &  SCIRD\\
				\hline
				F1 & $0.8189$ & $0.8204$  &  $0.8208$\\
				\hline
				
			\end{tabular}
		\end{center}
	\end{table}

	\subsection{Ablation Study: Impact of Regularizers and Multiscale Representation}
	\label{sec:variants}
	We report results for variants of our DRIS to demonstrate the benefit from each novel element in the network. We name the variants as follows: \color{black} 1) DRIS-Fixed - the representation layer comprises fixed SCIRD filters that undergo no further optimization; \color{black}2) DRIS - network with no regularizers; 3) DRIS-O, network with only the orientation diversity regularizer; 4) DRIS-N, network with only the noise robustness regularizer; and 4) DRIS-GP, network with both geometrical priors incorporated. The multiscale versions contain an MS prefix to the above notation. Table \ref{tab:results_var} reports the F1 and ACC values on the DRIVE and STARE datasets. It can be observed that MS-DRIS-GP performs the best as would be expected. Both the regularizers improve performance over (the fixed and unconstrained representation layer scenarios) DRIS-Fixed and DRIS with the noise robustness regularizer offering the larger relative advantage.  
	
\begin{figure*}
		\begin{center}
			\includegraphics[scale=.10]{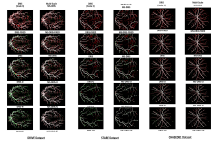}
		\end{center}
		\caption{\small{Visual comparison of different variants of our approach. \color{black}DRIS-Fixed and MS-DRIS-Fixed (First row) use fixed representation layers comprising SCIRD filters; DRIS and MS-DRIS results in the second row of the figure do {\em not} employ any regularizers and hence optimize the representation layer in an unconstrained manner\color{black}. White-TP, Green-FN, Red-FP, Black-TN}. The multi-scale version that employs both the orientation and noise regularizers (MS-DRIS-GP) achieves the best results with particularly enhanced accuracy in detecting thin vessels. These outputs correspond to the input and ground truth images shown in Fig. \ref{fig:sampImages}.}
		\label{fig:visual_illus}
		
	\end{figure*}

	Fig. \ref{fig:visual_illus} compares different variants of our proposal by showing the final segmentation map as achieved by each variant on a representative image from each of DRIVE and STARE datasets\footnote{This figure is best viewed in color. The actual retinal images and their groundtruth segmentation maps are in Fig.\ \ref{fig:sampImages}.}. Two trends can be clearly observed - 1) DRIS-N results in fewer false-positives (red color) compared to DRIS and DRIS-O, reiterating the importance of noise regularization; 2) thin vessels are more accurately segmented in the multiscale version - \color{black}namely, false negatives (green color) are less visible in the output of MS-DRIS-GP as compared to DRIS-GP \color{black}.  For the remainder of the paper, we report results using the MS-DRIS-GP variant of our technique.
	
\begin{table}
		\caption{\small{Evaluation of variants of DRIS-GP}}
		\label{tab:results_var}
		\begin{center}
			\begin{tabular}{cccc}
				\hline\hline
				\textbf{Method} & \textbf{Database} & \textbf{\textit{F1}} & \textbf{\textit{Acc}}\\
				\hline
				\multirow{2}{*}{DRIS-Fixed} & DRIVE & $0.7987$ & $0.9437$ \\
				& STARE  & $0.8021$ & $0.9570$ \\
				& CHASEDB1  & $0.7957$ & $0.9554$ \\
				\multirow{2}{*}{MS-DRIS-Fixed} & DRIVE & $0.8017$ & $0.9451$  \\
				& STARE  & $0.8093$ & $0.9581$ \\
				& CHASEDB1  & $0.8003$ & $0.9577$ \\
				\hline
				\hline
				\multirow{2}{*}{DRIS} & DRIVE & $0.8021$ & $0.9479$ \\
				& STARE  & $0.8116$ & $0.9608$ \\
				& CHASEDB1  & $0.8011$ & $0.9596$ \\
				\multirow{2}{*}{MS-DRIS} & DRIVE & $0.8072$ & $0.9499$  \\
				& STARE  & $0.8177$ & $0.9627$ \\
				& CHASEDB1  & $0.8062$ & $0.9617$ \\
				\hline
				\multirow{2}{*}{DRIS-O} & DRIVE & $0.8052$ & $0.9491$ \\
				& STARE  & $0.8162$ & $0.9619$  \\
				& CHASEDB1  & $0.8051$ & $0.9609$ \\
				\multirow{2}{*}{MS-DRIS-O} & DRIVE & $0.8115$ & $0.9505$ \\
				& STARE  & $0.8260$ & $0.9638$   \\
				& CHASEDB1  & $0.8125$ & $0.9632$ \\
				\hline
				\multirow{2}{*}{DRIS-N} & DRIVE & $0.8101$ & $0.9518$   \\
				& STARE  & $0.8214$ & $0.9633$  \\
				& CHASEDB1  & $0.8110$ & $0.9626$ \\
				\multirow{2}{*}{MS-DRIS-N} & DRIVE & $0.8182$ & $0.9542$  \\
				& STARE  & $0.8322$ & $0.9659$  \\
				& CHASEDB1  & $0.8173$ & $0.9656$ \\
				\hline
				\multirow{2}{*}{DRIS-GP} & DRIVE & $0.8145$ & $0.9530$ \\
				& STARE  & $0.8262$ & $0.9652$  \\	
				& CHASEDB1  & $0.8152$ & $0.9643$ \\			
				\multirow{2}{*}{MS-DRIS-GP} & DRIVE & $\textbf{0.8220}$ & $\textbf{0.9563}$  \\
				& STARE  & $\textbf{0.8364}$ & $\textbf{0.9687}$  \\
				& CHASEDB1  & $\textbf{0.8211}$ & $\textbf{0.9672}$ \\
				\hline\hline
			\end{tabular}
		\end{center}
	\end{table}
	

	\subsection{Broad Evaluation on a Standard Test-Train Configuration}
	\label{sec:specConfig}
	We report an exhaustive evaluation on a widely used test-train configuration that is consistent with state of the art methods from the last decade \cite{liskowski2016segmenting,li2016cross,orlando2017discriminatively,yan2018joint, wang2019blood}. For DRIVE, a fixed split of 20 training images and 20 test images is used across all the methods. For STARE, a leave-one-out validation procedure is carried out for learning based methods: the training and test cycle is repeated 20 times with 19 images reserved for training, and the remaining one used for evaluation. For CHASEDB1, the first 20 images are used for training and the remaining 8 are used for evaluation. Table \ref{tab:result_consi} reports an extensive comparison of MS-DRIS-GP with several other methods. Note that most of these numbers are reported from their respective papers. For DRIU \cite{maninis2016deep}, results on the DRIVE dataset are reported via the soft outputs shared by the authors publicly; while for STARE and CHASEDB1, we reproduced their publicly shared model using the standard configuration. For ML-UNET\cite{zhang2018deep}, we report numbers on DRIVE and STARE from their paper, and for CHASEDB1, we reproduce their implementation for our standard configuration. The first eight methods in the table are unsupervised techniques, and the remaining are learning based, with the last eight being deep learning methods.  \color{black}Methods marked by an asterisk indicate results reported without FOV. For our method, we include results both with and without FOV.\color{black} 
	
\begin{figure}
		\begin{center}
			\includegraphics[scale=.14]{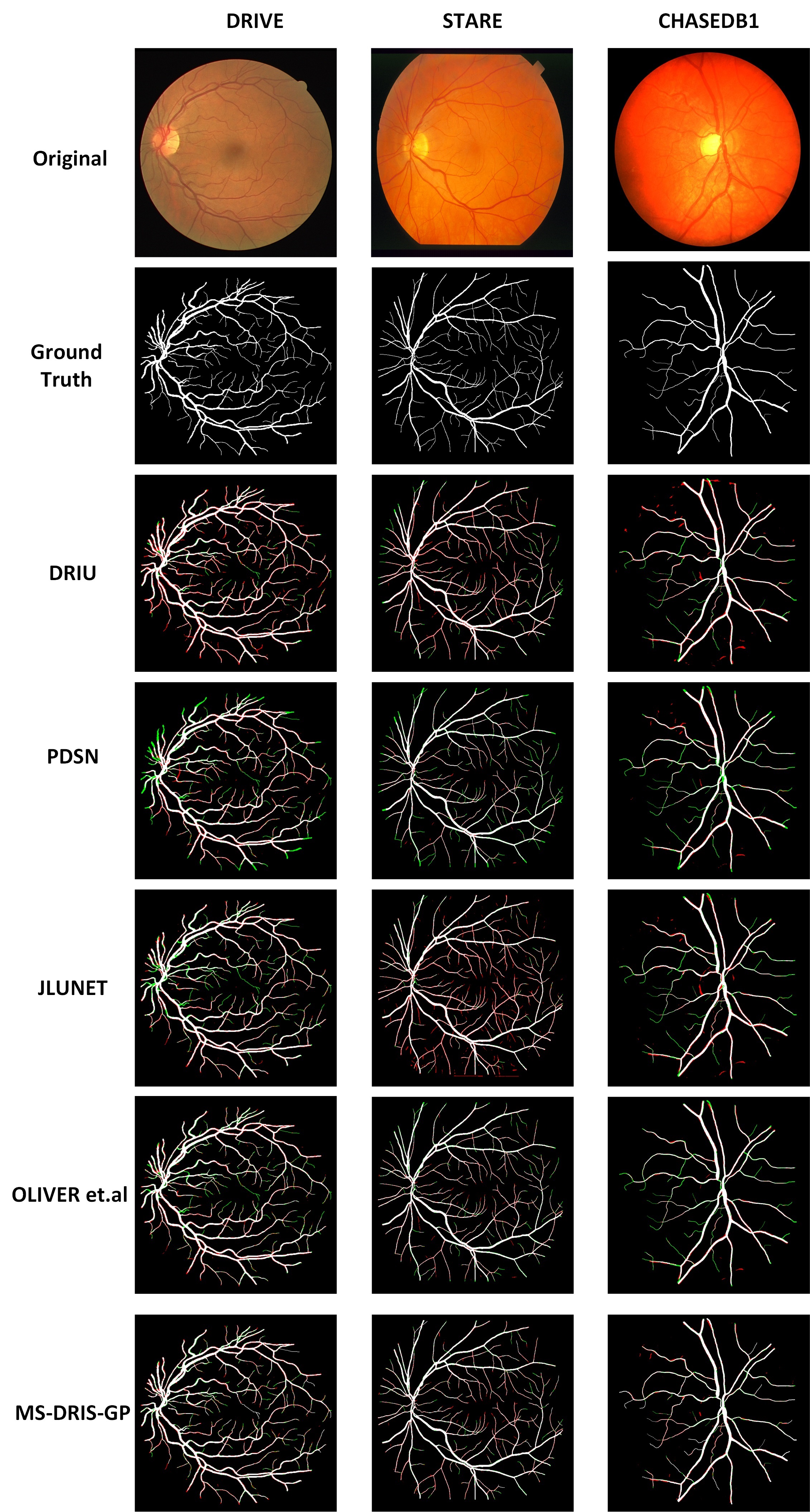}
		\end{center}
		\caption{\small{\color{black} Visual comparison against top competing deep learning methods. White-TP, Green-FN, Red-FP, Black-TN. MS-DRIS-GP achieves the best results with particularly enhanced accuracy in the detection of thin vessels\color{black}.}}
		\label{fig:visual_SOTR}
		
	\end{figure}

	MS-DRIS-GP yields compelling results on all three datasets. On DRIVE, ML-UNET \cite{zhang2018deep} produces better sensitivity measure while MS-DRIS-GP is best for all the other metrics. Similarly, on STARE, ML-UNET \cite{zhang2018deep} exhibits better performance by the specificity measure, but MS-DRIS-GP outperforms these techniques when assessed by the other metrics. On CHASE, MS-DRIS-GP produces best results for all the metrics. AUC is considered a particularly important measure for this problem and as Table \ref{tab:result_consi} confirms, MS-DRIS-GP produces the best AUC values on all 3 datasets. Interestingly, amongst the unsupervised methods, the results of Fan \etal\cite{FanTIP} are comparable to recent deep learning methods. \color{black} Visual comparisons against the top competing deep learning methods are shown in Fig. \ref{fig:visual_SOTR}. \color{black}
	
	\begin{table*}
		\caption{\small{Comparisons against competing methods on a standard train-test configuration. Asterisk symbol indicates that the results are without FOV. Our method has two rows - first row with FOV and second row without FOV}}
		\label{tab:result_consi}
		\begin{tabular}{l|l|llll|llll|llll}
			\hline
			\multicolumn{2}{c}{  } &
			\multicolumn{4}{c}{DRIVE} &
			\multicolumn{4}{c}{STARE} &
			\multicolumn{4}{c}{CHASEDB1} \\
			\textbf{Methods} & \textbf{Year} & \textbf{Sens} & \textbf{Spec} & \textbf{Acc} & \textbf{Auc} & \textbf{Sens} & \textbf{Spec} & \textbf{Acc} & \textbf{Auc}& \textbf{Sens} & \textbf{Spec} & \textbf{Acc} & \textbf{Auc}\\
			\hline
			2nd Human Observer & - & $0.7760$ & $0.9724$ & $0.9472$ & - & $0.8952$ & $0.9384$ & $0.9349$ & - &$0.8105$ & $0.9711$ & $0.9545$ & - \\
			\hline
			\multicolumn{14}{c}{Unsupervised methods}\\
			\hline
			Zhang \cite{zhang2010retinal} & 2010 & $0.7120$ & $0.9724$ & $0.9382$ & - & $0.7177$ & $0.9753$ & $0.9484$ & - &- & -& -& - \\
			You \cite{you2011segmentation} & 2011 & $0.7410$ & $0.9751$ & $0.9434$ & - & $0.7260$ & $0.9756$ & $0.9497$ & - &- & - & - & - \\
			Fraz \cite{fraz2012approach} & 2012 & $0.7152$ & $0.9759$ & $0.9430$ & - & $0.7311$ & $0.9680$ & $0.9442$ & - &- & - & - & - \\
			Roy \cite{roychowdhury2015iterative} & 2015 & $0.7395$ & $0.9782$ & $0.9494$ & $0.9672$ & $0.7317$ & $0.9842$ & $0.9560$ & $0.9673$ &$0.7615$ & $0.9575$ & $0.9467$ & $0.9623$ \\
			Azzopardi \cite{azzopardi2015trainable} & 2015 & $0.7655$ & $0.9704$ & $0.9442$ & $0.9614$ & $0.7716$ & $0.9701$ & $0.9497$ & $0.9563$ &$0.7585$ & $0.9587$ & $0.9387$ & $0.9487$ \\
			Yin \cite{yin2015vessel} & 2015 & $0.7246$ & $0.9790$ & $0.9403$ & - & $0.8541$ & $0.9419$ & $0.9325$ & - &- & - & - & - \\
			\color{black}Zhao \cite{zhao2015automated} & 2015 & $0.742$ & $0.982$ & $0.954$ & $0.862$ & $0.780$ & $0.978$ & $0.956$ & $0.874$ &- & - & - & - \\ \color{black}
			Zhang \cite{zhang2016robust} & 2016 & $0.7743$ & $0.9725$ & $0.9476$ & $0.9636$ & $0.7791$ & $0.9758$ & $0.9554$ & $0.9748$ &$0.7626$ & $0.9661$ & $0.9452$ & $0.9606$ \\
			Fan* \cite{FanTIP} & 2019 & $0.736$ & $0.981$ & $0.960$ & - & $0.791$ & $0.970$ & $0.957$ & - &$0.657$ & $0.973$ & $0.951$ & - \\
			\hline
			\hline
			\multicolumn{14}{c}{Supervised methods with hand-crafted features}\\
			\hline
			Marin \cite{marin2011new} & 2011 & $0.7067$ & $0.9801$ & $0.9452$ & $0.9588$ & $0.6944$ & $0.9819$ & $0.9526$ & $0.9769$ &- & - & - & - \\
			Fraz \cite{fraz2012ensemble} & 2012 & $0.7406$ & $0.9807$ & $0.9480$ & $0.9747$ & $0.7548$ & $0.9763$ & $0.9534$ & $0.9768$ &$0.7224$ & $0.9711$ & $0.9469$ & $0.9712$ \\
			\color{black} Wang \cite{wang2019retinal} & 2019 & - & - & $0.9547$ & - & - & - & $0.9646$ & - &- & - & $0.9617$ & - \\ \color{black}
			Wang \cite{wang2019blood} & 2019 & $0.7648$ & $0.9817$ & $0.9541$ & - & $0.7523$ & $0.9885$ & $0.9603$ & - &$0.7730$ & $0.9792$ & $0.9603$ & - \\
			\hline
			\multicolumn{14}{c}{Deep learning methods}\\
			\hline
			Li \cite{li2016cross} & 2016 & $0.7569$ & $0.9816$  & $0.9527$ & $0.9738$ & $0.7726$ & $0.9844$ & $0.9628$ & $0.9879$ &$0.7507$ & $0.9793$ & $0.9581$ & $0.9716$ \\
			
			\color{black} Fu* \cite{fu2016deepvessel} & 2016 & $0.7603$ & - & $0.9523$ & - & $0.7412$ & - & $0.9489$ & - &$0.7130$ & - & $0.9489$ & - \\ \color{black}
			Orlando \cite{orlando2017discriminatively} & 2017 & $0.7897$ & $0.9684$ & - & - & $0.7680$ & $0.9738$ & - & - &$0.7277$ & $0.9715$ & - & - \\
			Dasgupta* \cite{dasgupta2017fully} & 2017 & $0.7691$ & $0.9801$ & $0.9533$ & $0.9744$ & - & - & - & - &- & - & - & - \\
			PDSN \cite{liskowski2016segmenting} & 2016 & $0.7811$ & $0.9807$ & $0.9535$ & $0.9790$ & $0.8554$ & $0.9862$ & $0.9729$ & $0.9928$ & $0.7816$ & $0.9836$ & $0.9628$ & $0.9823$ \\
			DRIU \cite{maninis2016deep} & 2016 & $0.8280$ & $0.9728$ & $0.9541$ & $0.9801$ & $0.7919$ & $0.9827$ & $0.9706$ & $0.9814$ & $0.7651$ & $0.9822$ & $0.9657$ & $0.9746$ \\
			\color{black} Oliver \cite{oliveira2018retinal} & 2018 & $0.8039$ & $0.9804$ & $0.9576$ & $0.9821$ & $0.8315$ & $0.9858$ & $0.9694$ & $0.9905$ & $0.7779$ & $0.9864$ & $0.9653$ & $0.9855$ \\ \color{black}
			ML-UNET* \cite{zhang2018deep} & 2018 & $\textbf{0.8723}$ & $0.9618$ & $0.9504$ & $0.9799$ & $0.7673$ & $\textbf{0.9901}$ & $0.9712$ & $0.9882$ & $0.7667$ & $0.9825$ & $0.9649$ & $0.9839$ \\
			JL-UNET \cite{yan2018joint} & 2018 & $0.7653$ & $0.9818$ & $0.9542$ & $0.9752$ & $0.7581$ & $0.9846$ & $0.9612$ & $0.9801$ & $0.7633$ & $0.9809$ & $0.9610$ & $0.9781$ \\
			\color{black} Gu* \cite{gu2019net} & 2019 & $0.8309$ & - & $0.9545$ & $0.9779$ & - & - & - & - & - & - & - & - \\\color{black}
			\multirow{2}{*}{MS-DRIS-GP (Ours)}& - & $0.8426$ & $\textbf{0.9823}$ & $\textbf{0.9603}$ & $\textbf{0.9844}$ & $\textbf{0.8667}$ & $0.9871$ & $\textbf{0.9734}$ & $\textbf{0.9930}$ & $\textbf{0.8025}$ & $\textbf{0.9874}$ & $\textbf{0.9693}$ & $\textbf{0.9858}$ \\
			& - & $0.8425$ & $\textbf{0.9849}$ & $\textbf{0.9723}$ & $\textbf{0.9870}$ & $\textbf{0.8664}$ & $0.9895$ & $\textbf{0.9803}$ & $\textbf{0.9935}$ & $\textbf{0.8017}$ & $\textbf{0.9908}$ & $\textbf{0.9788}$ & $\textbf{0.9864}$ \\
		\end{tabular}
	\end{table*}
	
	\begin{table*}[tbh]
		\caption{\small{Focused comparisons against selected state-of-art methods.}}
		\label{tab:results_comp}
		\begin{center}
			\begin{tabular}{l|lll|lll|lll}
				\hline
				\multicolumn{1}{c}{   } &
				\multicolumn{3}{c}{DRIVE} &
				\multicolumn{3}{c}{STARE} &
				\multicolumn{3}{c}{CHASEDB1} \\
				\textbf{Methods} &  \textbf{F1} & \textbf{Acc} & \textbf{Auc} & \textbf{F1}  & \textbf{Acc} & \textbf{Auc} & \textbf{F1}&  \textbf{Acc} & \textbf{Auc}\\
				\hline
				Fan \cite{FanTIP}  & $0.7914$ & $0.9485$ & $-$ & $0.8021$ &  $0.9516$ & $-$ & $0.7951$  & $0.9481$ & $-$ \\
				PDSN \cite{liskowski2016segmenting} & $0.7950$  & $0.9531$ & $0.9711$ & $0.8132$ &  $0.9584$ & $0.9804$ & $0.8043$ & $0.9447$ & $0.9663$ \\
				DRIU \cite{maninis2016deep}  & $0.7929$ & $0.9428$ & $0.9703$ & $0.8164$  & $0.9566$ & $0.9842$ & $0.8019$ & $0.9619$ & $0.9635$ \\
				ML-UNET \cite{zhang2018deep}  & $0.7986$ & $0.9437$ & $0.9718$ & $0.8116$ & $0.9553$ & $0.9825$ & $0.7997$  & $0.9594$ & $0.9634$ \\
				JL-UNET \cite{yan2018joint}  & $0.8040$ & $0.9549$ & $0.9721$ & $0.8227$ &  $0.9631$ & $0.9836$ & $0.8086$  & $0.9621$ & $0.9674$ \\
				\color{black} Oliver \etal \cite{oliveira2018retinal}  & $0.8148$ & $0.9552$ & $0.9796$ & $0.8270$ &  $0.9659$ & $0.9861$ & $0.8132$  & $0.9658$ & $0.9771$ \\ \color{black}
				MS-DRIS-GP (Ours)& $\textbf{0.8220}$  & $\textbf{0.9563}$ & $\textbf{0.9814}$ & $\textbf{0.8364}$ & $\textbf{0.9687}$ & $\textbf{0.9903}$ & $\textbf{0.8211}$ & $\textbf{0.9672}$ & $\textbf{0.9833}$ \\

			\end{tabular}
		\end{center}
	\end{table*}
	
	\subsection{Focused Comparisons against State of The Art Methods}
	\label{sec:results}
\begin{figure}
		\begin{center}
			\includegraphics[scale=.14]{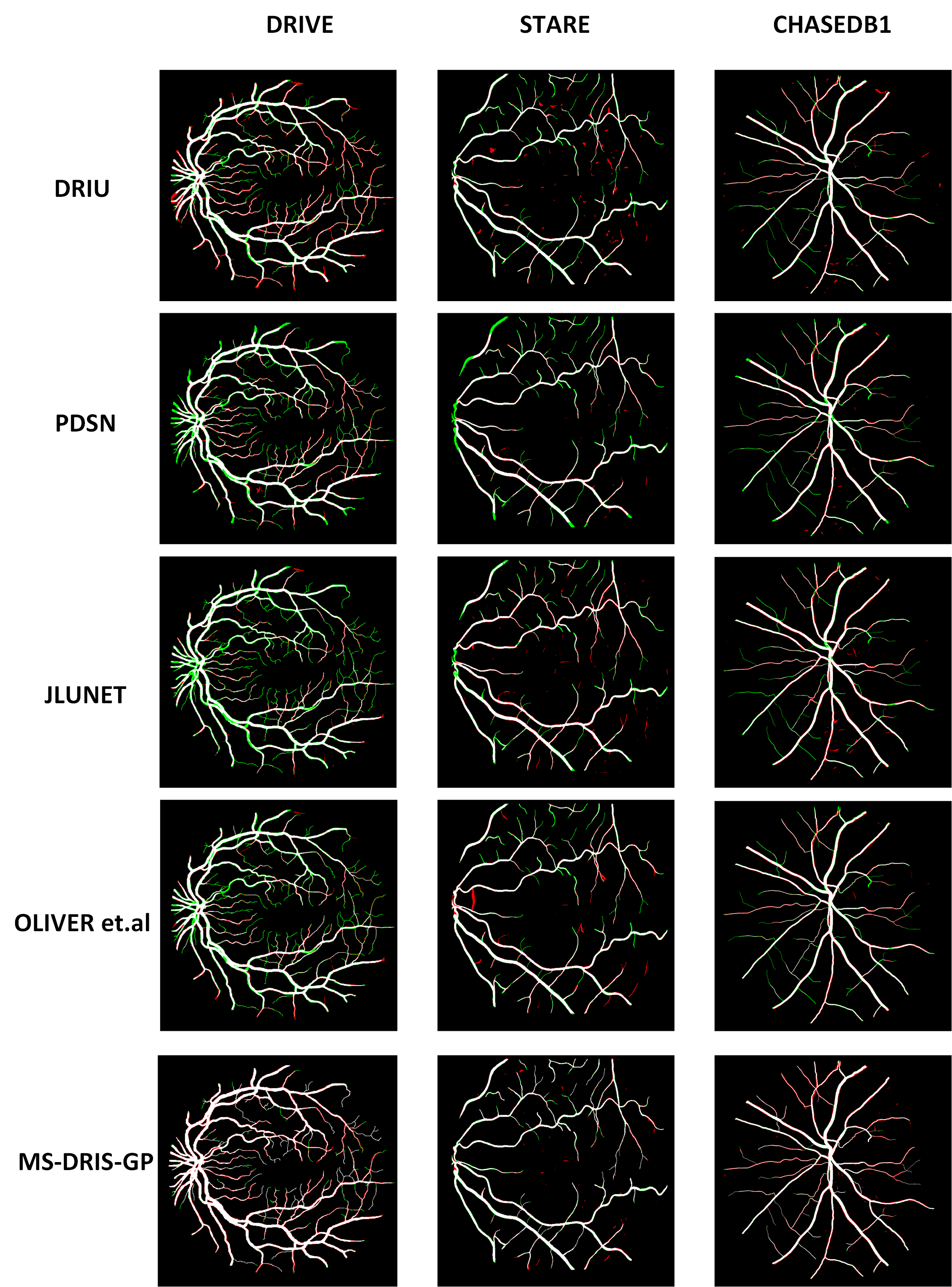}
		\end{center}
		\caption{\small{\color{black} Visual comparison against the deep learning methods in Table \ref{tab:results_comp}. White-TP, Green-FN, Red-FP, Black-TN. MS-DRIS-GP achieves the best results with particularly enhanced accuracy in the segmentation of thin vessels. These outputs correspond to the input and ground truth images shown in Fig. \ref{fig:sampImages}.\color{black}}}
		\label{fig:visual_SOTR_SP}
		
	\end{figure}

	The standard train-test configuration for the results reported in Table \ref{tab:result_consi} suffers from selection bias for the DRIVE and CHASEDB1 datasets. Further, the train-test configuration employed for STARE is a relatively imbalanced setup. To mitigate these issues, we report results on our configuration described in Section \ref{sec:expsetup}. Table \ref{tab:results_comp} reports F1, ACC and AUC values against the following recent state-of-the art methods that yielded the most promising results in Table \ref{tab:result_consi}: 
	
	\begin{itemize}
		\item PDSN (TMI'2016) \cite{liskowski2016segmenting} - classifies each pixel separately by considering a patch around it, hence we call it Pixel level Deep Segmentation Network (PDSN). 
		\item DRIU (MICCAI'2016) \cite{maninis2016deep} - uses VGGNET \cite{simonyan2014very}, fine tuned to segment retinal vessels and the optic disk. A class-balancing entropy loss function is used to learn the network parameters. 
		\item ML-UNET (MICCAI'2018) \cite{zhang2018deep} - a very recent deep method employing a U-net architecture with multiple labels for distinguishing between thin and thick vessels; we call it Multi-Label- Unet (ML-UNET). 
		\item JL-UNET (TBME'2018) \cite{yan2018joint} - another recent deep learning method that employs a segment level loss function jointly with a pixel-level loss function on a Unet architecture; we call it Joint Loss-Unet (JL-UNET).
		\color{black}\item Oliver \etal \cite{oliveira2018retinal} (Expert Sys. 2018) - Another recent deep learning approach that uses a Multi-Scale Stationary Wavelet transform to pre-process the retinal images, followed by data-augmentation operations such as rotation prior to training a standard U-NET like architecture. \color{black}
		\item Fan \etal \cite{FanTIP} (To appear: TIP'2019) - is a very recent unsupervised method that integrates a hierarchical strategy into image matting model for blood vessel segmentation.  
	\end{itemize}
	Note that averaged results over different test-train configurations is not directly applicable for Fan \etal \cite{FanTIP} as it is not a learning based method. Hence, in Table \ref{tab:results_comp}, we report the average results on all the images available for this method. JL-UNET and ML-UNET tackle the issue of thin vs thick vessels, and are hence interpretable as multi-scale methods. Thresholds to calculate F1 values for each method were chosen as suggested in their respective papers\footnote{We gratefully acknowledge DRIU and JL-UNET authors for providing us code/output images; and we faithfully reproduced the PDSN and ML-UNET implementations by confirming that our implementation produces results that are fully consistent with those reported in their papers.}.	
	We note from Table \ref{tab:results_comp} that MS-DRIS-GP outperforms state of the art for all three metrics on the 3 datasets. \color{black} Visual comparisons are illustrated in Fig. \ref{fig:visual_SOTR_SP}. As can be observed, MS-DRIS-GP detects thin vessels more accurately with reduced instances of false positives. \color{black}
	
\begin{figure*}
		\begin{center}
			\includegraphics[scale=.25]{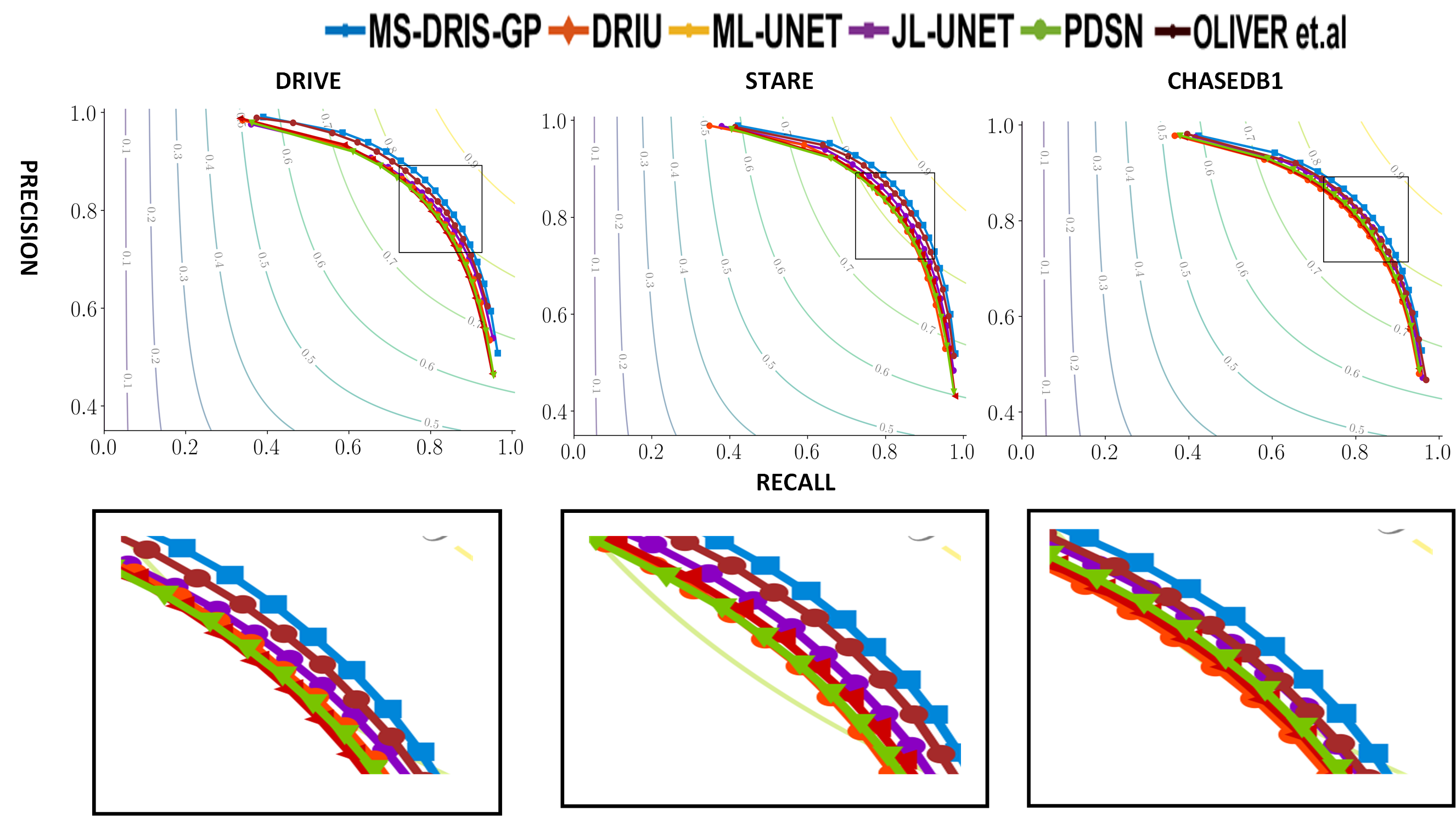}
		\end{center}
		\caption{\small{Precision-recall (PR) curves over the 3 datasets. Top row represents the complete curves. Bottom row represents the zoomed portion of the top row.}}
		\label{fig:prec-recall}
	\end{figure*}

	Another significant evaluation methodology for binary segmentation problems is to analyze the trade-off between precision and recall, defined as\footnote{Note that the AUC measure and precision-recall (PR) curves are not reported for Fan \etal\cite{FanTIP} -- this is also the case in their paper -- because their threshold selection strategy is different from the threshold used on the soft output of deep learning methods.}: 
	$Prec = \frac{TP}{TP+FP}\mbox{,  }Recall\mbox{ }(sens) = \frac{TP}{TP + FN}$. Since all the deep learning methods produce a continuous output, for consistency, we calculated the values of precision and recall for various segmentation thresholds and the corresponding Precision-Recall (PR) curves are shown in Fig. \ref{fig:prec-recall}. The optimum points on the curve for all the methods are zoomed in for greater clarity. As can be observed, MS-DRIS-GP achieves a superior PR curve for all the 3 datasets.  

	\begin{figure}
		\begin{center}
			\includegraphics[scale=.25]{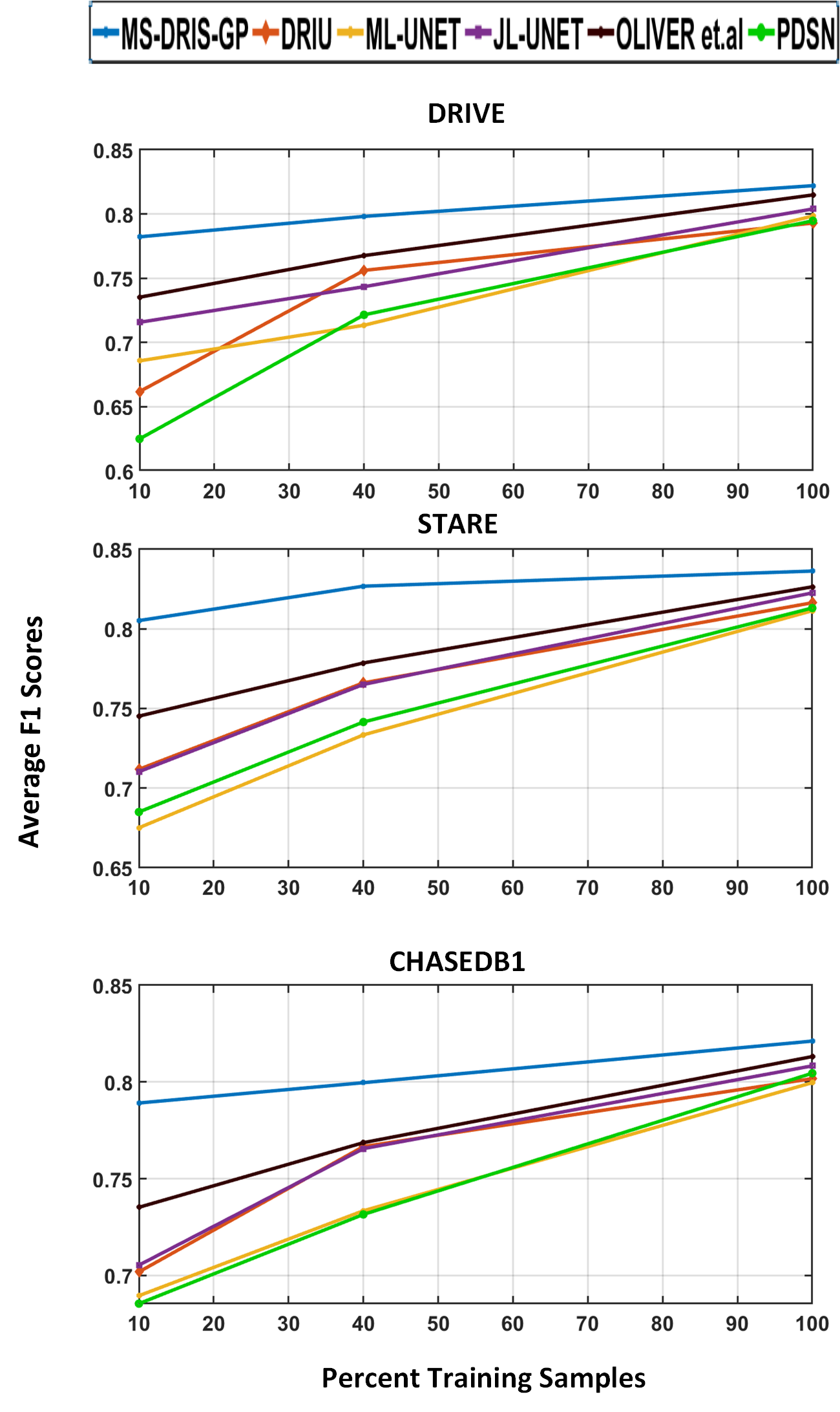}
		\end{center}
		\caption{\small{F1 vs Percent training samples}}
		\label{fig:lowtrain}
		
	\end{figure}

	\subsection{Performance under Various Training Regimes}
	\label{sec:lowtrain}
	
	To evaluate the robustness of our approach to reduced training, we train the top competing deep networks respectively with 40 and 10 percent of training samples used in previous experiment. A plot of Averaged F1 scores of 5 random selections as a function of training set size for all the 3 datasets is shown in Fig. \ref{fig:lowtrain}. MS-DRIS-GP exhibits a graceful degradation with a decrease in the number of training images, and outperforms all the competing methods by a significant margin especially for the case of 10 percent training. 
	
\begin{figure}
		\centering
		\begin{center}
			\includegraphics[scale=.25]{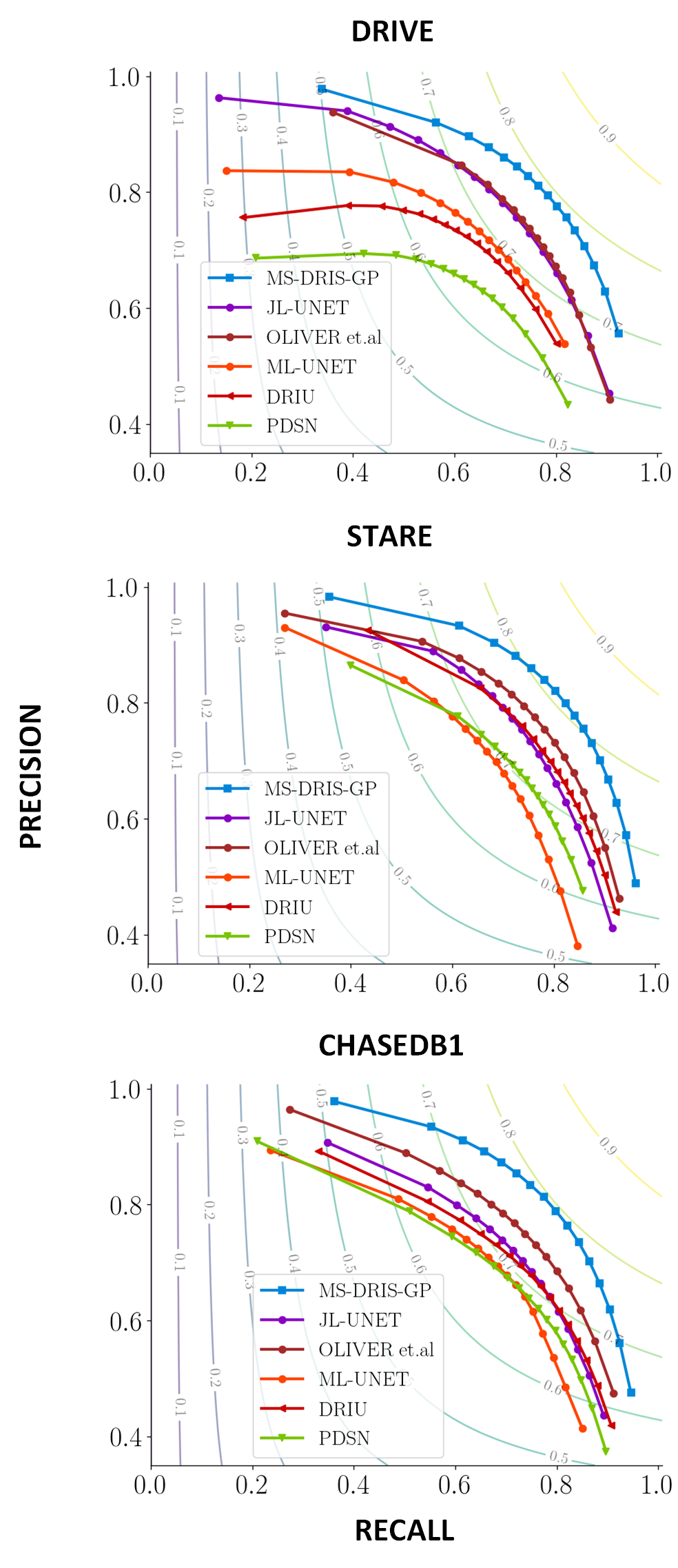}
		\end{center}
		\caption{\small{Precision-recall (PR) curves for the 10 $\%$ training setup.}}
		\label{fig:low_prec}
	\end{figure}

	Fig. \ref{fig:low_prec} is a plot of the PR curves for the 10$\%$ training scenario for all the 3 datasets. We observe a much wider margin between MS-DRIS-GP and  state of the art deep learning methods. This is to be expected because MS-DRIS-GP derives its model partially from meaningful priors, which play a critical role in scenarios where training is limited \cite{shen2017deep, srinivas2015structured}. This benefit is not available to purely data driven methods.

	\color{black}
	\subsection{Evaluation on High-Resolution Fundus Images (HRF)}
	\label{sec:HRF}
	To further validate the effectiveness and robustness of our method, we perform an experiment on the HRF dataset \cite{odstrcilik2013retinal, zhao2017automatic} comprising 45 high-resolution images divided into healthy, diabetic retinopathy and glaucomatous sets. Every set comprises 15 fundus images, each of size $3504 \times 2336$ pixels. The first 5 images of every set are used for training and the remainder are used for evaluation, a protocol consistent with other methods evaluated on this data-set \cite{orlando2017discriminatively, yan2018joint}. Sample images from each set along with their ground-truth segmentation maps and the results obtained by our method are illustrated in Fig. \ref{fig:HRF}. Similar to other methods \cite{orlando2017discriminatively, yan2018joint}, to reduce the computational burden on our network, we downsample the images and labels by a factor of 2 and train our network. Testing is performed on downsampled images and upsampled for evaluation. Note that the size of the downsampled images is $1752\times 1168$ which is almost twice the size of the other datasets evaluated in this work. Hence, the vessel structures are also assumed to be scaled accordingly in these images. The size of representation layer filters is increased by 4 across each dimension, resulting in sizes: $7\times 7$, $9\times 9$, $11\times 11$, $13\times 13$ and $15\times 15$. The sizes of the other parameters described in Section \ref{sec:multi-scale} are scaled accordingly. Table \ref{tab:results_HRF} reports the comparisons against other state-of-the-art methods that have been previously evaluated on this dataset. The numbers are taken from the respective references. As can be observed, MS-DRIS-GP excels in all metrics by a significant margin thereby validating the benefits of our approach. The results and our model for the HRF dataset are also shared at: \url{https://scholarsphere.psu.edu/concern/generic_works/mcv43nz236}
	
		\begin{figure}
		\centering
		\begin{center}
			\includegraphics[scale=.20]{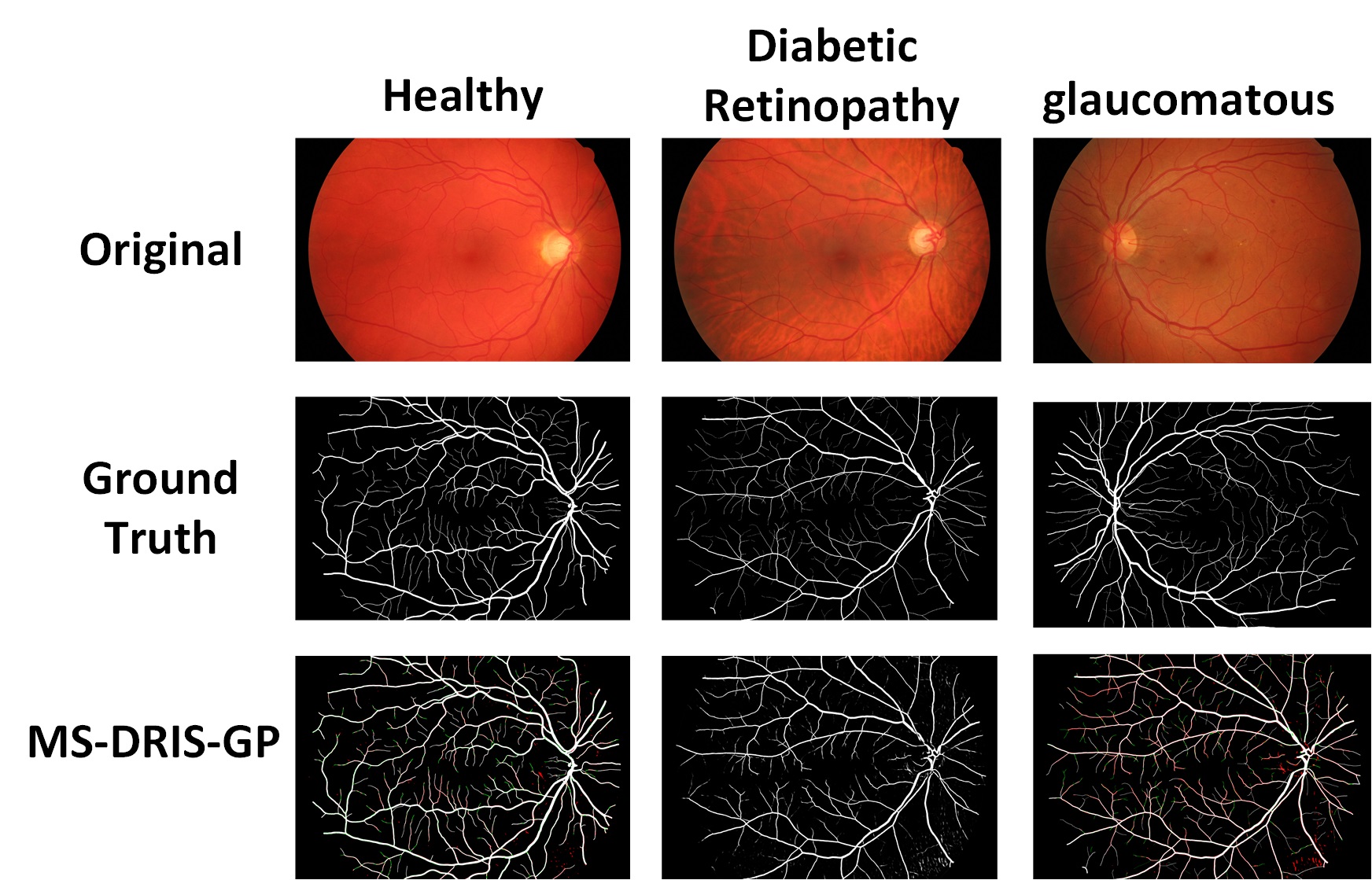}
		\end{center}
		\caption{\small{Sample images from HRF dataset.}}
		\label{fig:HRF}
	\end{figure}
	
		\begin{table}
			\color{black}
		\caption{\small{Comparisons on HRF dataset}}
		\label{tab:results_HRF}
		\begin{center}
			\begin{tabular}{ccccc}
				\hline\hline
				\textbf{Method} & \textbf{\textit{Sens}}& \textbf{\textit{Spec}}& \textbf{\textit{Acc}} & \textbf{\textit{F1}}\\
				\hline
				Orlando \cite{orlando2017discriminatively}& $0.7874$ & $0.9584$ & - & $ 0.7199$ \\
				\hline
				JL-UNET \cite{yan2018joint}& $0.8084$ & $0.9417$ & $0.9437$ & $0.7211$\\
				\hline
				Wang \cite{wang2019retinal}& - & - & $0.9573$ & $0.7474$\\
				\hline
				MS-DRIS-GP (ours) & $\textbf{0.8144}$ & $\textbf{0.9733}$ & $\textbf{0.9588}$  & $\textbf{0.7832}$\\
				\hline\hline
			\end{tabular}
		\end{center}
	\color{black}
	\end{table}
	
	\color{black}
	\subsection{Cross Training Evaluation}
	\label{sec:crosstrain}
	In this section we test the ability of MS-DRIS-GP to generalize across different datasets. Cross training studies are conducted in a number of other methods \cite{yan2018joint, li2016cross}, wherein the network is trained on one dataset and evaluated on a different dataset. Table \ref{tab:results_cross} reports a comparison of MS-DRIS-GP against these methods. As can be observed, MS-DRIS-GP outperforms the competing methods under all the evaluation metrics, demonstrating that incorporation of domain priors improves generalization capability. 
	
	\color{black}
	\subsection{Segmentation of Thin Vessels} 
	
	Thin vessels can provide crucial indications for retinopathy and are particularly challenging to segment \cite{almotiri2018retinal}.
	To illustrate the effectiveness of our method in detecting thin vessels, we show a comparison with JL-UNET \cite{yan2018joint} and ML-UNET \cite{zhang2018deep} which are specifically designed for this purpose and are shown to be the state-of-the-art methods for detecting thin vessels. Vessels with thickness less than 3 pixels are denoted as thin vessels and rest of the vessels are denoted as thick vessels in their work. 
	
	Focusing on ground truth comprising only thin vessels (as is done in \cite{yan2018joint}), we report key quantitative metrics in Table \ref{tab:results_thin}. Note that to be consistent with similar experiments in \cite{yan2018joint}, \cite{zhang2018deep}; the results in Table \ref{tab:results_thin} are for the DRIVE dataset and the train-test configuration in Section \ref{sec:specConfig}.
	Three standard metrics -- namely Specificity, Precision and AUC -- have been used \cite{yan2018joint}, \cite{zhang2018deep} for evaluation of thin vessel segmentation. Consistent with \cite{yan2018joint}, to calculate the evaluation metrics, each thin
	vessel (in the ground truth) is assigned with a 5-pixel searching range and pixels in a given output segmentation map located within the range are counted for pixel-to-pixel matching.
	
	Table \ref{tab:results_thin} confirms that MS-DRIS-GP performs better than ML-UNET and JL-UNET in all the metrics and hence excels at detecting thin vessel structures. 
	\color{black}
	
	\begin{table}
		\caption{\small{Comparisons for cross-training}}
		\label{tab:results_cross}
		\begin{center}
			\resizebox{\linewidth}{!}{
				\begin{tabular}{cccccc}
					\hline\hline
					\textbf{Dataset} & \textbf{Method} & \textbf{\textit{Sens}} & \textbf{\textit{Spec}}& \textbf{\textit{Acc}}& \textbf{\textit{Auc}}\\
					\hline
					\multirow{2}{*}{DRIVE (Trained on STARE)} & Li \cite{li2016cross} & $0.7273$ & $0.9810$ & $0.9486$ & $0.9677$ \\ 
					& JL-UNET \cite{yan2018joint} & $0.7292$ & $0.9815$ & $0.9494$ & $0.9599$ \\
					& MS-DRIS-GP (ours) & $\textbf{0.7723}$ & $\textbf{0.9830}$ & $\textbf{0.9560}$ & $\textbf{0.9769}$  \\
					\hline
					\multirow{2}{*}{STARE (Trained on DRIVE)} & Li \cite{li2016cross} & $0.7027$ & $0.9828$ & $0.9545$ & $0.9671$ \\ 
					& JL-UNET \cite{yan2018joint}& $0.7211$ & $0.9840$ & $0.9569$ & $0.9708$ \\
					& MS-DRIS-GP (ours) & $\textbf{0.7778}$ & $\textbf{0.9864}$ & $\textbf{0.9708}$ & $\textbf{0.9818}$  \\
					\hline\hline
			\end{tabular}}
		\end{center}
	\end{table}
	
	\begin{table}
		\caption{\small{Comparisons for thin vessels}}
		\label{tab:results_thin}
		\begin{center}
			\begin{tabular}{cccc}
				\hline\hline
				\textbf{Method} & \textbf{\textit{Spec}}& \textbf{\textit{Prec}}& \textbf{\textit{AUC}}\\
				\hline
				ML-UNET \cite{zhang2018deep}& $0.9003$ & $0.7202$ & $0.8678$ \\
				\hline
				JL-UNET \cite{yan2018joint}& $0.9158$ & $0.7449$ & $0.8948$ \\
				\hline
				MS-DRIS-GP (ours) & $\textbf{0.9314}$ & $\textbf{0.7534}$ & $\textbf{0.9036}$  \\
				\hline\hline
			\end{tabular}
		\end{center}
	\end{table}
	
	\subsection{Network Parameters and Inference Time}
	\label{sec:computation}
	
	Table \ref{tab:results_memory} compares the number of network parameters and inference times of MS-DRIS-GP with competing deep learning methods. The number of parameters for our MS-DRIS-GP is estimated as follows: The representation network as described in Section \ref{sec:multi-scale} has 5 filter banks with 12 filters in each bank constituting $3420$ parameters. The bridge connection between representation and task network is a $1\times 1\times 60\times 32$ filter which constitutes $1920$ parameters. The task network has $14$ layers with $32$ $3\times 3\times 32$ filters in each layer resulting in $129,024$ parameters. The final layer is a $3\times 3\times 32$ filter with 288 learnable parameters. The total number of parameters is $134,652 \approx 1.3\times 10^{5} $. The number of parameters for competing methods as reported in Table \ref{tab:results_memory} is computed in a similar fashion, wherein the architectures are derived from their respective papers. For all methods, the inference times are reported on an NVIDIA Titan X GPU (12GB). 
	
	It is readily observed that MS-DRIS-GP has the fewest parameters (less than 10$\%$ of state of the art) and the smallest inference time among its competitors. This result also corroborates the intuition that incorporating domain knowledge leads to a design that is memory and computation efficient. PDSN's inference time is particularly high because a deep network is employed to classify each pixel individually rather than segmenting the image as a whole.

	\begin{table}[t]
		\caption{\small{Network Size and Inference time comparison. Inference time is averaged over test images obtained from the DRIVE dataset.}}
		\label{tab:results_memory}
		\begin{center}
			\begin{tabular}{l|l|l}
				\hline
				\textbf{Methods} &  \textbf{Network Parameters} & \textbf{Inf. Time}\\
				\hline
				PDSN \cite{liskowski2016segmenting} & $\approx 1.1 \times 10^{6} $  & 92s \\
				DRIU \cite{maninis2016deep}  & $\approx 4.6 \times 10^{6}$ & 34ms \\
				ML-UNET \cite{zhang2018deep} & $\approx 4.3 \times 10^{6} $  & 33ms \\
				JL-UNET \cite{yan2018joint}  & $\approx 1.8 \times 10^{7}$ & 40ms \\
				MS-DRIS-GP (Ours)& $\approx \mathbf{1.3 \times 10^{5}}$  & $\textbf{26ms}$\\

			\end{tabular}
		\end{center}
	\end{table}

	\section{Summary and Conclusion}
	\label{sec:summ}
	We present a new deep learning paradigm for retinal image segmentation that parses the overall network into representation and task layers/network, incorporates trainable domain-specific priors into the representation layer, and jointly optimizes representation and task parameters. The form of the priors takes inspiration from hand-crafted features, and encourages the representation filters to respond to curvilinear vessel structures while ignoring domain-specific noise. The approach defines a new state-of-art performance under a wide variety of experimental settings and evaluation metrics. Future work includes conceiving additional means to incorporate domain knowledge into deep retinal segmentation networks. One idea along this vein would be to incorporate structural priors on the network outputs that capture vessel structures of higher order. \color{black}Further, these ideas can be incorporated in other curvilinear segmentation problems such as lung vessel detection in CT scans \cite{gu2019net}, wrinkle detection  on face images \cite{batool2015fast}, road detection \cite{kong2010general} in aerial images, etc. \color{black} 
	
	\bibliographystyle{IEEEtran}
	\bibliography{refs}
\end{document}